\documentclass[%
 twocolumn,
 reprint,
showpacs,
 amsmath,amssymb,
 aps,
]{revtex4}

\usepackage{color}
\usepackage{ulem}
\usepackage{graphicx}
\usepackage{dcolumn}
\usepackage{bm}

\begin{document}

\preprint{APS/123-QED\author{Takahiro Ohgoe}}

\title{Ground-State Phase Diagram of the Two-Dimensional Extended Bose-Hubbard Model}

\author{Takahiro Ohgoe$^1$}
\author{Takafumi Suzuki$^2$}
\author{Naoki Kawashima$^1$}%
\affiliation{%
	$^1$Institute for Solid State Physics, University of Tokyo, Kashiwa, Chiba 277-8581, Japan\\
	$^2$Research Center for Nano-Micro Structure Science and Engineering, Graduate School of Engineering, University of Hyogo, Himeji, Hyogo 671-2280, Japan
}%

\date{\today}

\begin{abstract}
We investigate the ground-state phase diagram of the soft-core Bose-Hubbard model with the nearest-neighbor repulsion on a square lattice by using an unbiased quantum Monte Carlo method. In contrast to the previous study[P. Sengupta {\it et. al.}, Phys. Rev. Lett. {\bf 94}, 207202 (2005)], we present the ground-state phase diagrams up to large hopping parameters. As a result, in addition to the known supersolid above half-filling, we find supersolid even below and at half-filling for large hopping parameters. Furthermore, for the strong nearest-neighbor repulsion, we show that the supersolid phase occupies a remarkably broad region in the phase diagram. The results are in qualitative agreement with that obtained by the Gutzwiller mean-field approximation[M. Iskin, Phys. Rev. A {\bf 83}, 051606(R) (2011) and T. Kimura, Phys. Rev. A {\bf 84}, 063630 (2011)].

\end{abstract}

\maketitle


\section{\label{sec1}Introduction}
	Supersolid has attracted great interest for a long time as a fascinating quantum state that has superfluidity and solidity simultaneously. In the early theoretical works by Andreev and  Lifshitz\cite{andreev1969}, and by Chester\cite{chester1970}, they proposed a scenario that supersolid might appear when zero-point defects in solid such as $^{4}$He undergo Bose-Einstein condensate at low temperatures without destroying the crystal structure. After several decades, a discovery was made in 2004 by Kim and Chan\cite{kim2004-1, kim2004-2}. In their experiments on solid $^{4}$He, they observed nonclassical rotational inertia associated with superfluidity in the solid. After the discovery, further theoretical or experimental works\cite{prokofev2005, boninsegni2006, rittner2006} provided the evidence that it is different from a bulk supersolid of the Andreev-Lifshits-Chester scenario. The superfluidity in solid $^{4}$He seems to appear due to the extended defects such as grain-boundaries\cite{sasaki2006, pollet2007} or dislocations\cite{boninsegni2007}.

	In contrast to the supersolid in continuous spaces, supersolid in lattice systems has been a promising candidate recently. This is based on the recent experimental development of optical lattice systems\cite{greiner2002, sage2005, ni2008, ospelkaus2008}. Ultra-cold Bose gases trapped in optical lattice are ideal systems to realize the Bose-Hubbard models\cite{jaksch1998}. From the intensive theoretical and numerical studies, the existence of supersolid phases has been established in the extended Bose-Hubbard models\cite{otterlo1995, batrouni2000, goral2002, wessel2005, boninsegni2005, sengupta2005, batrouni2006, yi2007, suzuki2007, dang2008, yamamoto2009, danshita2009, pollet2010, capogrosso2010-1, xi2011, yamamoto2012}. Most of the supersolids in lattice systems are achieved by doping particles or holes into insulating solid states at commensurate filling factors. If doped defects delocalize and Bose-Einstein condensate against a phase separation, supersolid appears by the the Andreev-Lifshits-Chester scenario. Thus, resulting supersolids are stabilized at incommensurate filling factors.

	One of the simplest models to study supersolids is the soft-core Bose-Hubbard model with nearest-neighbor repulsions. By accurate quantum Monte Carlo calculations on this model, checherboard supersolid phases have been found on a 1D chain\cite{batrouni2006}, a 2D square lattice\cite{sengupta2005}, and a 3D simple cubic lattice\cite{yamamoto2009, xi2011, ohgoe2012}. In the 1D and 2D cases, supersolid regions are found only above half-filling (interstitial supersolid). In contrast, in the 3D case, supersolids are also found even below and at half-filling for large hopping parameters (vacancy supersolid and commensurate supersolid respectively)\cite{yamamoto2009, xi2011, ohgoe2012}. Especially, the presence of supersolid at the commensurate filling factor 1/2 is fascinating as an exceptional supersolid {\it without any doping}, although such supersolid regions have not been found so far in 1D and 2D. Therefore, it is a question why there is a discrepancy between 2D and 3D systems.
	
	Recent works based on the Gutzwiller mean-field approximation have provided some interesting results on the ground-state phase diagram of the model\cite{kimura2011,iskin2011}, including a possible answer to the above question. In the ground-state phase diagram presented in Ref. \cite{kimura2011}, the author found a supersolid phase below and at half-filling. Since it was found more clearly for larger hopping parameters, he suggested that the absence of such supersolid regions in the 2D quantum Monte Carlo study\cite{sengupta2005} might be due to the not so large hopping parameter. Since the regions for the supersolid below and at half-filling are much smaller than that above half-filling and the mean-field approximations tend to overestimate the region of the supersolid phase\cite{yamamoto2009, ohgoe2012}, the existence of such supersolid regions is a subtle problem. As discussed in Ref. \cite{kimura2011}, more precise treatments are desirable to conclude the existence of 2D supersolid phase below and at half-filling, because the Gutzwiller approximation becomes more accurate in higher dimensions and particle densities.

	The other interesting result presented in Ref. \cite{iskin2011} is on supersolid phases for strong nearest-neighbor repulsion. The ground-state phase diagrams show that, as the nearest-neighbor repulsion increases, the supersolid phase expands up to large hopping parameters in the phase diagram. Especially, the 2D case of this result might be most important, because it has the possibility of realizations in quasi-2D dipolar Bose gases whose the dipoles are polarized along the $z$-axis\cite{griesmaier2005}. Therefore, from the viewpoint of experiments, we also need to determine the more precise phase boundaries in the 2D system and check the accuracy of the phase diagram.  
	
	In this paper, motivated by the results of the Gutzwiller treatment, we investigate the ground-state phase diagram of the extended Bose-Hubbard model on a square lattice by numerically exact quantum Monte Carlo simulations. The paper is organized as follows. In Sec. \ref{sec2}, we describe the model discussed in this papers and the quantum Monte Carlo method we used. Sec. \ref{sec3} presents the ground-state phase diagrams in the grand canonical ensembles. These phase diagrams include up to the third insulating lobes. Within this region, we confirm that our ground-state phase diagrams are qualitative agreement with those obtained by the Gutzwiller approximation. In Sec. \ref{sec4}, we study quantum phase transitions and explain the procedure of determining the phase boundaries presented in the previous section. In Sec. \ref{sec5}, we investigate the supersolid phase at half-filling by obtaining results for the canonical ensembles. By showing a ground-state phase diagram at half-filling, we confirm that the supersolid phase is easily found for large hopping parameters. Finally, in Sec. \ref{sec6}, we summarize our results.
  
\section{\label{sec2}MODEL AND METHOD}
	The model considered in this paper is the soft-core Bose-Hubbard model with nearest-neighbor repulsions on a square lattice. The Hamiltonian is given by
	\begin{eqnarray}
		H & = & - t \sum_{\langle i, j\rangle} ( b_{i}^{\dagger} b_{j} + h.c. ) - \mu \sum_{i} n_{i} + \frac{U}{2} \sum_{i} n_{i} (n_{i}-1) \nonumber \\ & &+ V \sum_{\langle i,j \rangle} n_i n_j.
	\end{eqnarray}
	Here, $b^{\dagger}_{i}$($b_{i}$) is the bosonic creation (annihilation) operator on site $i$, and $n_{i}$ is the particle number operator defined as $n_{i} = b^{\dagger}_{i} b_{i}$. The summation $\langle i, j \rangle$ is taken over all pairs of nearest-neighbor sites. For a square lattice, the coordination number $z$ equals 4. Furthermore, $t$ is the hopping parameter, $\mu$ is the the chemical potential, $U$ is the on-site interaction, and $V$ is the nearest-neighbor interaction. In this paper, we consider the case where the interactions are repulsive ($U, V>0$). In our simulations, we treat $N = L \times L$ systems with the periodic boundary condition. 
	
	In the classical limit $t/U=0$, the ground-states are known and simple\cite{sengupta2005, iskin2011, xi2011}. When the nearest-neighbor repulsion satisfies $zV/U < 1$, the ground states are checkerboard solids at filling factors $\rho=1/2$, 3/2,..., and uniform Mott-insulators at $\rho=1$, 2, ... . To characterize each state, we can label it as $(n_A, n_B)$ which represents a pair of particle numbers on the two sublattices $A$ and $B$. Without loss of generality, we assume that $n_A\geq n_B$. Based on this notation, the ground states are labeled as (1,0), (1,1), (2,1), (2,2), ... at $\rho=1/2$, 1, 3/2, 2, ... respectively. In contrast, for $zV/U > 1$, all ground states are checkerboard solids. The states are labeled as (1,0), (2,0), (3,0), (4,0), ... at $\rho=1/2$, 1, 3/2, 2, ... respectively, and the transition from $\rho=n/2$ to $(n+1)/2$ takes place at $(\mu/U)_c=n$, when the chemical potential is increased. Therefore, $zV/U=1$ is a critical point for $\rho \geq 1$ in the classical limit. When the finite $t/U$ is introduced, the critical point $(zV/U)_c=1$ is shifted to slightly larger values due to quantum fluctuation.
	
	To investigate the properties of the model for finite values of $t/U$, we used an unbiased quantum Monte Carlo method. The formulation we used is based on the Feynman path integral representation. In the representation, the $d$-dimensional quantum system is mapped to the ($d+1$)-dimensional classical systems. In the mapped systems, each configuration is called world-line with $d$-dimensional space axises and one-dimensional imaginary time axis. Based on this representation, we sample the world-line configurations according to the Markov chain Monte Carlo. To update the configurations, we used the worm-type algorithm\cite{prokofev1998, syljuasen2002, kawashima2004, kato2009}.

\section{\label{sec3}GROUND-STATE PHASE DIAGRAM IN THE GRAND-CANONICAL ENSEMBLE}
	In this section, we present ground-state phase diagrams in the $zt/U$-$\mu/U$ plane. The recent Gutzwiller mean-field study suggested that the supersolid phase might exist even below half-filling for large hopping parameters\cite{kimura2011}. In addition, the other work provided the results that the ground-state phase diagram have qualitatively different structures between weak nearest-neighbor repulsions and strong nearest-neighbor repulsions\cite{iskin2011}. Remarkably, in the latter case, the supersolid phase seems to occupy very large region in the phase diagram. To confirm these results by numerically exact quantum Monte Carlo calculations, we show the ground-state phase diagrams at $zV/U=1$ and $zV/U=1.5$ in Secs. \ref{subsec3-1} and \ref{subsec3-2} respectively.

	\subsection{\label{subsec3-1}Ground-state phase diagram at $zV/U=1$}
	
	\begin{figure}[h]
		\includegraphics[width=9cm]{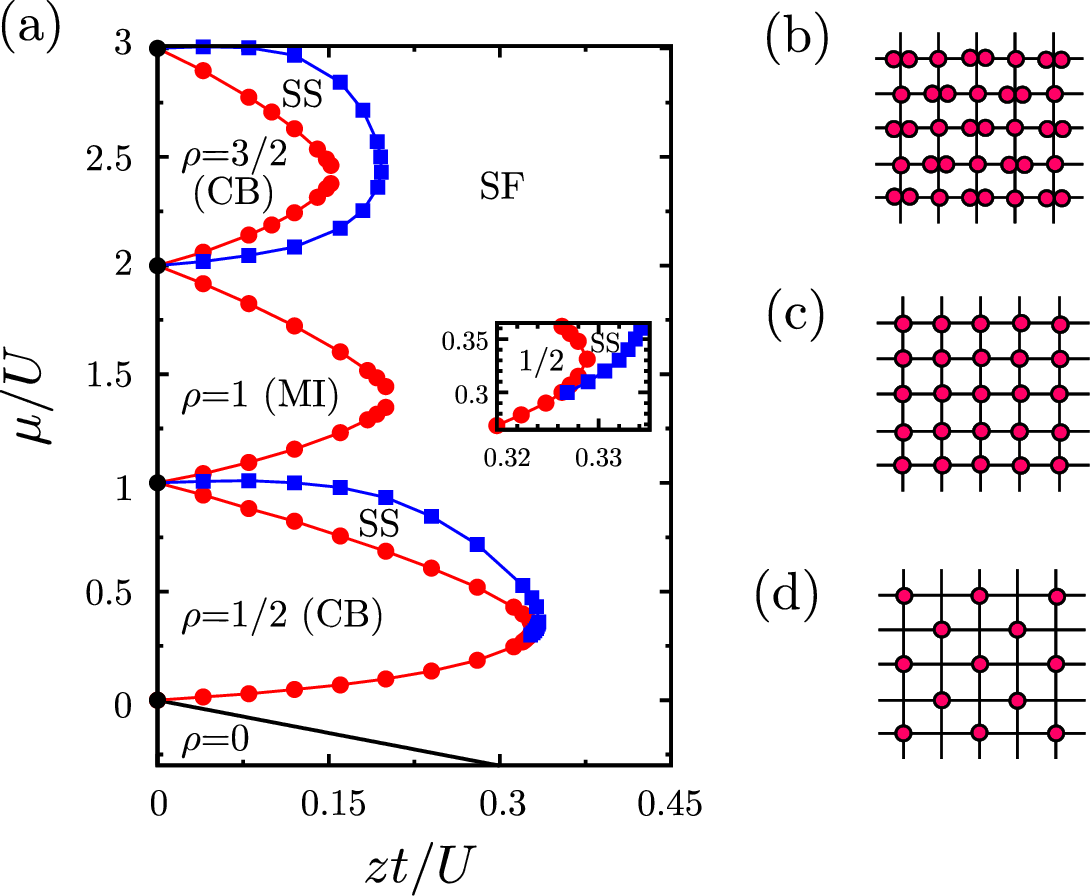}
		\caption{\label{fig:phasediag1} (Color online) (a) Ground-state phase diagram of the extended Bose-Hubbard model on a square lattice at $zV/U=1$. Red circles indicate boundaries for the insulating lobes. Blue squares represent the SS-SF boundary. The inset is the enlarged view of the region around the tip of the first CB lobe. Error bars are drawn but most of them are much smaller than the symbol size (here and in the following figures). Black line is the boundary between the empty region and the SF that can be obtained analytically. Other lines are used to guide the eyes. (b), (c), and (b) Schematic configurations for the insulators at $\rho=3/2$, $\rho=1$, and $\rho=1/2$ respectively. Each red circle represents one particle on the sites.}
	\end{figure}

    \begin{figure}[h]
		\includegraphics[width=9cm]{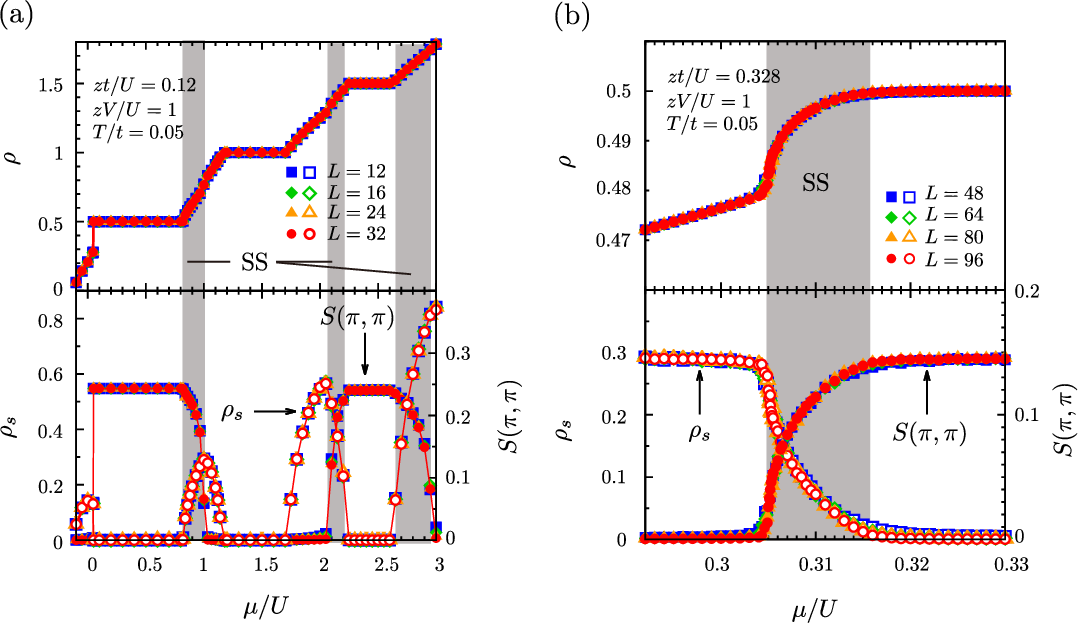}
		\caption{\label{fig:quantitiesV0.25} (Color online) (a) and (b) Physical quantities as functions of $\mu/U$ at $(zt/U, zV/U, T/t)=(0.12, 1, 0.05)$ and $(zt/U, zV/U, T/t)=(0.328, 1, 0.05)$ respectively. Shaded regions indicate the supersolid state where $\rho_s$ and $S(\pi, \pi)$ take finite values simultaneously.}
	\end{figure}

	In Fig. \ref{fig:phasediag1} (a), we show the ground-state phase diagram at $zV/U=1$ in the $zt/U$-$\mu/U$ plane. To detect each phase, we measured the particle density $\rho = 1/N \sum_{i} \langle n_{i} \rangle$, the superfluid stiffness $\rho_{s} = \langle \mbox{\boldmath $W$}^2 \rangle/(2dt\beta L^{d-2})$, and the structure factor $S( \mbox{\boldmath $k$} ) = 1/N^2 \sum_{i, j} e^{i \mbox{\boldmath $k$} \cdot \mbox{\boldmath $r$}_{ij}} (\langle n_{i} n_{j} \rangle - \langle n_{i} \rangle^2 )$. Here, $\langle \cdots \rangle$ is the thermal average, and $\mbox{\boldmath $W$}$ denotes the winding number vector in the path integral representation\cite{pollock1987}. $\beta$ represents the inverse temperature defined by $\beta=1/T$, $d$ is the dimensionality of system that is equal to 2 in this paper, $\mbox{\boldmath $k$}$ is the wave vector, and  $\mbox{\boldmath $r$}_{ij}$ indicates the relative position vector between sites $i$ and $j$. In our phase diagram up to $\mu/U \leq 3$, in addition to a conventional superfluid phase(SF), there are three insulating lobes at $\rho=1/2$, $\rho=1$, and $\rho=3/2$. Schematic configurations are shown in the Fig. \ref{fig:phasediag1} (b), (c), and (d), respectively. The lobe at $\rho=1$ is a uniform Mott-insulating phase (MI), and the others at $\rho=1/2$ and $\rho=3/2$ are checkerboard-type solid phases (CB) characterized by finite value of $S(\pi, \pi)$. We also confirm the presence of supersolid phases (SS) around the insulating CB lobes. The determinations of the phase boundaries are explained in detail in Sec. \ref{sec4}.

	To show the existence of each phase, we plot $\mu/U$ dependence of the measured quantities at $(zt/U, zV/U, T/t)=(0.12, 1, 0.05)$ and (0.328, 1, 0.05) in Fig. \ref{fig:quantitiesV0.25} (a) and (b) respectively. In the case of the small hopping parameter $zt/U=0.12$ in Fig. \ref{fig:quantitiesV0.25} (a), SS phases exit above $\rho=1/2$ and around $\rho=3/2$. When particles are removed from the checkerboard solid at $\rho=1/2$, possible supersolid is unstable against a phase separation as known by strong-coupling argument\cite{sengupta2005}. In contrast, for the larger hopping parameter $zt/U = 0.328$ in Fig. \ref{fig:quantitiesV0.25} (b), we find that SS phase are present even below half-filling. As seen in the inset of Fig. \ref{fig:phasediag1} (a), the SS phase covers the tip of the first CB lobe. This result suggests that the supersolid can be also stabilized at half-filling. In Sec. \ref{sec5}, we present direct evidence for supersolid at half-filling by obtaining results for the canonical ensemble and excluding possible phase separations. In addition to the SS around $\rho=1/2$, the other SS phase around $\rho=3/2$ more clearly covers the tip of the corresponding insulating CB lobe. Therefore, the supersolid seems to be stabilized even at $\rho=3/2$. The present 2D ground-state phase diagram is in qualitative agreement with that in 3D\cite{ohgoe2012} and that obtained by the Gutzwiller approximation\cite{otterlo1995, kovrizhin2005, iskin2011}. However, we find that the supersolid regions clearly become smaller as the dimensionality decreases.

	\subsection{\label{subsec3-2}Ground-state phase diagram at $zV/U=1.5$}
\begin{figure}[h]
		\includegraphics[width=9cm]{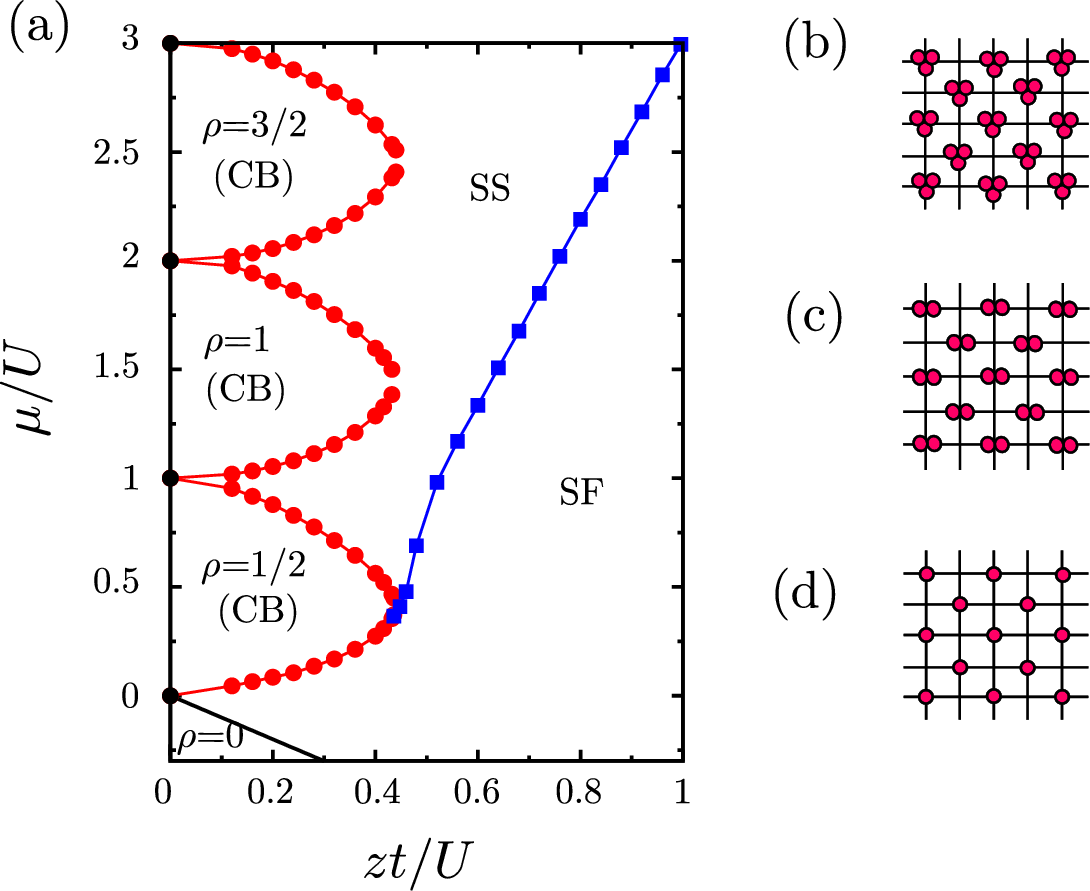}
		\caption{\label{fig:phasediag2} (Color online) Ground-state phase diagram in the $zt/U-\mu/U$ plane at $zV/U=1.5$. (b), (c) and (d) Schematic configurations for the insulators at $\rho=3/2$, $\rho=1$, and $\rho=1/2$ respectively. }
	\end{figure}

		For strong nearest-neighbor repulsions, all insulating states are checkerboard solid states and, thus, the ground-state phase diagram are quite different from that for weak nearest-neighbor repulsions. In Fig. \ref{fig:phasediag2} (a), we present the ground-state phase diagram at $zV/U=1.5$ in the ground-canonical ensemble. In contrast to the phase diagram at $zV/U=1$, all three insulating Mott lobes are actually the checkerboard solid ones. The schematic configurations at $\rho=3/2$, 1, and 1/2 are shown in Fig. \ref{fig:phasediag2} (b), (c), and (d) respectively. Compared with the case of $zV/U=1$, the insulating lobes extend up to larger hopping parameters. This result is reasonable, because the strong nearest-neighbor repulsion favors the checkerboard solid state. The remarkable point is that the connected SS phase exits, surrounding all the CB lobes. The SS phase occupies a broad region up to large hopping parameters, and the phase boundary behaves linearly. Our result is still in qualitative agreement with that obtained by the Gutzwiller approximation\cite{iskin2011}. However, the supersolid region is apparently smaller.

\begin{figure}[h]
		\includegraphics[width=9.5cm]{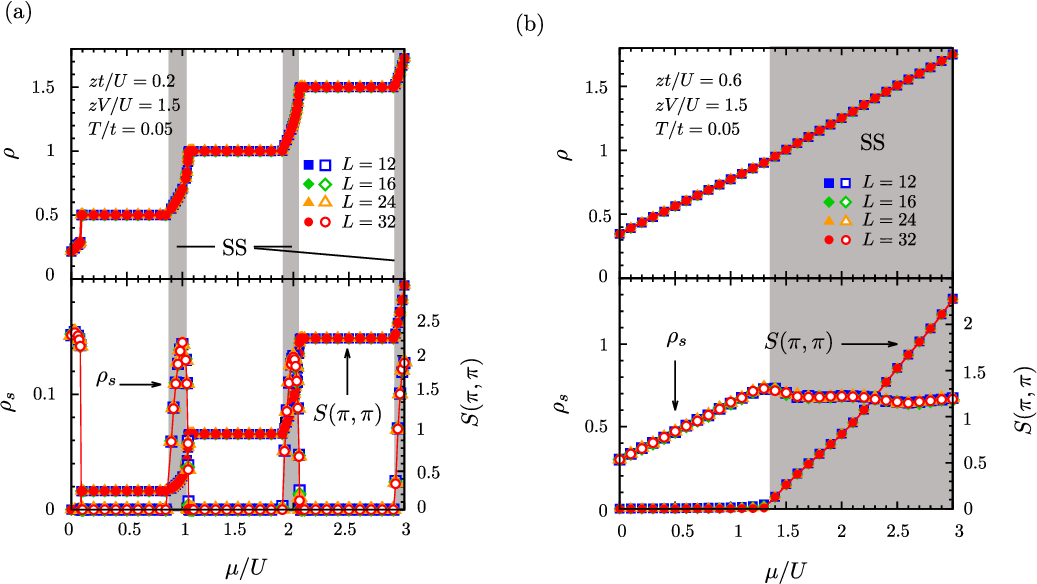}
		\caption{\label{fig:quantitiesV0.375} (Color online) (a) and (b) Physical quantities as functions of $\mu/U$ at $(zt/U, zV/U, T/t)=(0.2, 1.5, 0.05)$ and $(zt/U, zV/U, T/t)=(0.6, 1.5, 0.05)$ respectively.}
	\end{figure}

		To support the results, we plot the measured quantities as functions of $\mu/U$ at $(zt/U, zV/U, T/t)=(0.2, 1.5, 0.05)$ and (0.6, 0.15, 0.05) in Fig. \ref{fig:quantitiesV0.375} (a) and (b) respectively. In Fig. \ref{fig:quantitiesV0.375} (a), there are three plateaus at $\rho=1/2, 1, 3/2$, where $S(\pi, \pi)$ takes finite value. These plateaus correspond to the CB phases. Between these regions, $S(\pi, \pi)$ and $\rho_s$ take finite value simultaneously, indicating the SS phase. In contrast, just below $\rho=1/2$, there is no SS phase and we observed a clear discontinuity in the particle density again. Just below $\rho=1$,and 3/2, the slopes in the particle density are very steep. However, compared with that below $\rho=1/2$, possible discontinuities are not so clear. Thus, the CB-SS transitions might be weakly-first-order or second-order at this parameter. When the hopping parameter becomes smaller, we confirmed that the slopes become steeper, suggesting the presence of a first-order transition predicted by the strong coupling arguments\cite{sengupta2005}. For larger hopping parameter as in Fig. \ref{fig:quantitiesV0.375} (b), all the insulating plateaus disappear. In contrast, the SS phases are connected and occupy all the region for large chemical potentials.

\section{\label{sec4}QUANTUM PHASE TRANSITIONS}
	In this section, we study quantum phase transitions and explain how the phase boundaries are determined. There are three different kinds of quantum phase transitions in terms of symmetry breaking: the transition between two phases with different broken symmetries (the CB-SF transition), the superfluid transition that involves the gauge symmetry (the CB-SS transition and the MI-SF transition), and the checkerboard-order transition where the translational symmetry is breaking (the SS-SF transition). Since these quantum phase transitions have different properties related to the broken symmetries, we need different treatments to determine the phase boundaries. In the following three subsections, we explain the treatments for each phase boundary.

	\subsection{\label{subsec4-1}Solid-superfluid transition}
		We begin with the CB-SF transition that appears at the lower boundary of the first CB lobe. As observed in Figs. \ref{fig:quantitiesV0.25} (a) and \ref{fig:quantitiesV0.375} (a) as well as the previous quantum Monte Carlo works\cite{sengupta2005, xi2011}, there are finite jumps in the particle density at the boundary, indicating a first-order transition. This result can be understood from an argument on the broken symmetries in each phase and the standard Landau-Ginzburg-Wilson paradigm. In the CB phase, the broken symmetry is the $Z_2$ associated to the broken translational symmetry. On the other hand, in the SF phase, the $U(1)$ gauge symmetry is broken {\it at zero temperature}. (Note that, at finite temperatures in two dimensions, the SF phase shows not the long-range order, but the quasi-long range order.) According to the Landau-Ginzburg-Wilson paradigm, a transition between two phases with different broken symmetries results in a first-order transition or intermediate region where both symmetries are broken simultaneously. Since an intermediate supersolid phase is absent at the boundary, the direct CB-SF transition should be a first-order. Thus, we simply determined the phase boundary from the position of the finite jump in the particle density. 
	
	\subsection{\label{subsec4-2}Solid-supersolid transition and Mott-insulator-superfluid transition}
		At the CB-SS boundaries and MI-SF boundaries, the quantum phase transitions are the insulator-superfluid ones. As for the value of the dynamical critical exponents $z_c$, two possibilities are expected: generic transition with $z_c=2$ and special transition with $z_c=1$\cite{fisher1989}. Because of the difference, we have to determine the transition points in different manners.

\begin{figure}[h]
		\includegraphics[width=6cm]{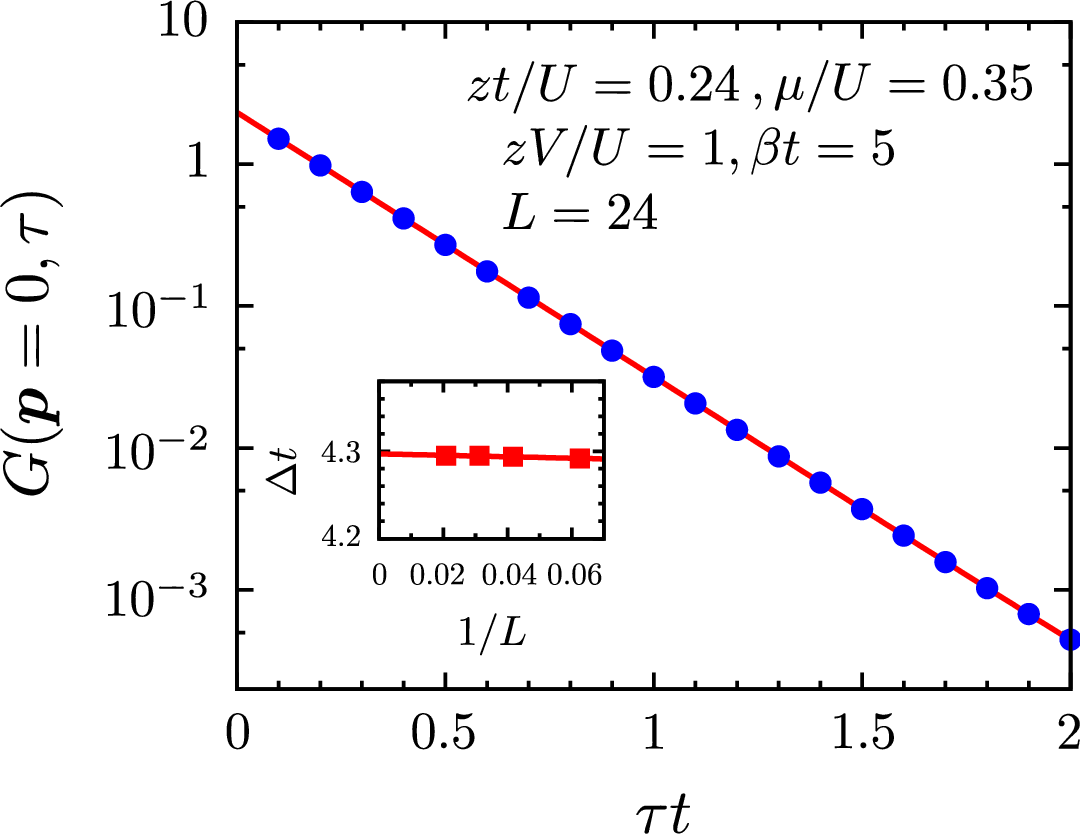}
		\caption{\label{fig:zero_mom_correl} (Color online) Extraction of the energy gap $\Delta_{(+)}$ from the zero-momentum Green function $G(\mbox{\boldmath $p$}=0, \tau)$ in the first CB lobe. Solid circles denotes the results obtained by our simulation, and the line represents the exponential fit. The inset shows the extrapolation of the obtained $\Delta$ (red squares) to the thermodynamic limit.}
	\end{figure}

		The generic transitions are driven by adding/removing a particle to/from the insulating phases. In this case, the phase boundary can be determined from the finite-size scaling analysis of $\rho_s$ for quantum critical points with $z_c=2$\cite{kato2010}. However, it can also be determined more simply from the the zero-momentum Green function $G(\mbox{\boldmath $p$}=0, \tau)$\cite{capogrosso2007, capogrosso2008}. In the worm algorithm, the zero-momentum Green function can be obtained by measuring the Matsubara Green function $G(\mbox{\boldmath $r$}_{i}, \tau)= \langle T_{\tau} b_i(\tau) b_0^{\dagger}(0) \rangle$. Here, $T_{\tau}$ indicates the time-ordering operator on the imaginary time $\tau$, and $b_i(\tau)$ is defined by $b_i(\tau)=e^{\tau H} b_i e^{-\tau H}$. From the asymptotic exponential decay $G(\mbox{\boldmath $p$}=0,\tau) \to Z_{+} e^{-\Delta_{+} \tau}$ ($\tau \to +\infty$) [$Z_{-} e^{\Delta_{-} \tau}$ ($\tau \to -\infty$)], we can estimate the energy gap $\Delta_{+}$ ($\Delta_{-}$) for creating single particle (hole) excitation with $\mbox{\boldmath $p$}=0$ in the insulating phases. In the ground-canonical ensemble, the energy gap corresponds to the distance between the observed point and the phase boundary in the $\mu$ direction. Thus, we determined the phase boundary from the energy gap. Fig. \ref{fig:zero_mom_correl} shows an example of estimating the energy gap $\Delta_{+}$ in the first CB lobe.
	
	\begin{figure}[h]
		\includegraphics[width=9cm]{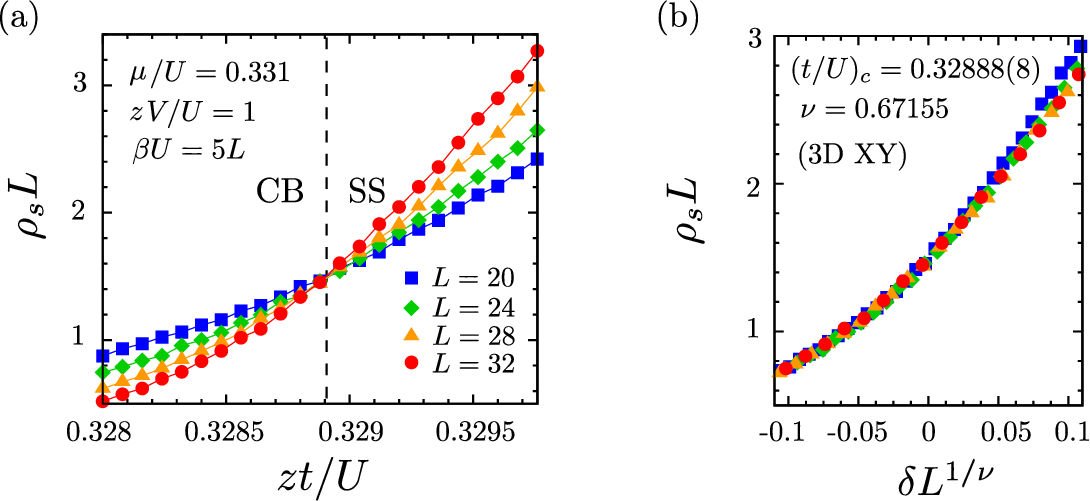}
		\caption{\label{fig:CBtoSS} (Color online) (a) Plots of $\rho_s L$ as functions of $t/U$ near the tip of the first CB lobe. The vertical dashed line is placed at the quantum critical point $(zt/U)_c=0.32888(8)$ that is estimated from the crossing point. (b) Scaling plots of $\rho_s L$. }
	\end{figure}
		In contrast to the generic transition, the special transition is driven by delocalizing quantum fluctuation. This transition occurs at the tip of insulating lobes with fixed $\mu/U$. The tip corresponds to a multicritical point where $z_c$ equals 1 due to a particle-hole symmetry\cite{fisher1989}. Therefore, to determine the critical point close to the tip in the inset of Fig. \ref{fig:phasediag1}, we performed the finite-size scaling analysis of $\rho_s$ for quantum phase transitions with $z_c=1$. In this analysis, the scaling form is given by $\rho_s L^{d+z_c-2} = f(\delta L^{1/\nu}, \beta/L^{z_d})$, where $\nu$ is the critical exponent of the correlation length, $\delta$ denotes the distance from critical points as $\delta=zt/U-(zt/U)_c$, and $f$ is a scaling function. In the present case of $d=2$ and $z_c=1$, the value of $d+z_c-2$ equals 1. Therefore, $\rho_s L$ should cross at the critical point for different system sizes with fixed $\beta/L$ and we can simply estimate it from the crossing point. Fig. \ref{fig:CBtoSS} (a) shows one example of this estimation. In this figure, we estimated the critical point as $(zt/U)_c=0.32888(8)$ for $\mu/U=0.331$ that is very close to the tip.

	To clarify the universality class of the special transition, we proceed to perform the finite-size scaling analysis of $\rho_s L$. In the case of $z_c=1$, the effective dimension becomes $d+z_c=3$. Since the breaking symmetry in this transition is related to the global $U(1)$ symmetry, this quantum phase transition is expected to belong to the 3D XY universality class. Using the critical exponent $\nu=0.67155$ of the 3D XY universality class \cite{campostrini2001} and the dynamical critical exponent $z_c=1$, we plot $\rho_s L$ as a function of $\delta L^{1/\nu}$ in Fig. \ref{fig:CBtoSS} (b). In the figure, we successfully observe the data collapse for large system sizes, supporting the validity of the present analysis.
	
	\subsection{\label{subsec4-3}Supersolid-superfluid transition}
\begin{figure}[h]
		\includegraphics[width=8cm]{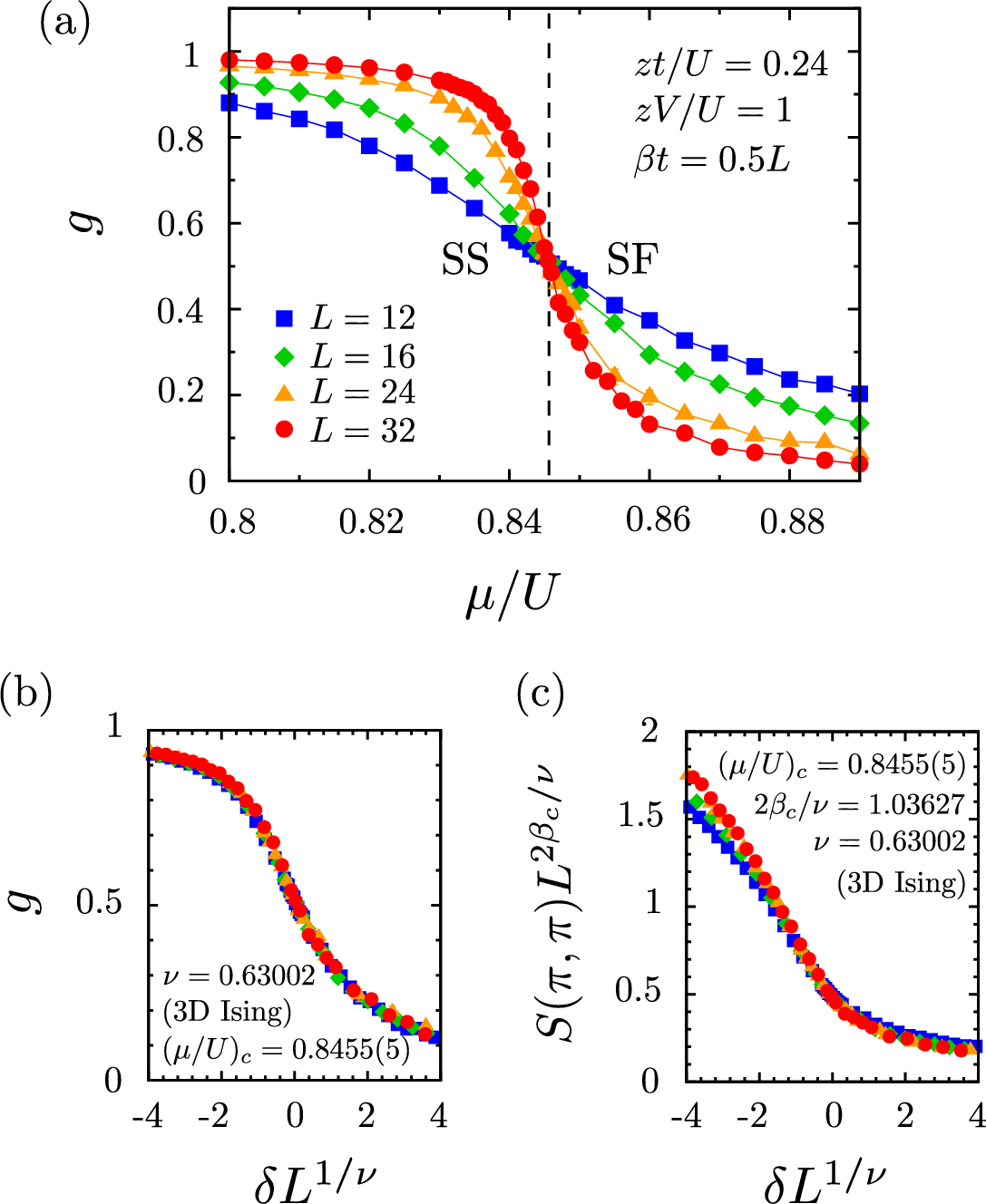}
		\caption{\label{fig:SStoSF} (Color online) (a) Estimation of the SS-SF boundary from a crossing point of the Binder ratio $g$ for different system sizes. (b) and (c) Finite-size scaling plots of $g$ and $S(\pi, \pi) L^{2\beta_c/\nu}$ respectively. }
	\end{figure}

		Finally, we explain the SS-SF boundaries. The SS-SF transition is the checkerboard-solid transition related to the $Z_2$ symmetry breaking of the translational symmetry. For this quantum phase transition, the critical point can be determined from the Binder ratio $g$ defined by $g=1/2[3-\langle m^4 \rangle/\langle m^2 \rangle^2]$ Here, $m$ indicates the order parameter defined by $m=1/N \sum_{i} n_{i} e^{i \mbox{\boldmath $k$} \cdot \mbox{\boldmath $r$}_{i}}$ with $\mbox{\boldmath $k$}=(\pi, \pi)$. The scaling form for $g$ is given by $g=f(\delta L^{1/\nu}, \beta/L^{z_c})$, where $\delta = zt/U -(zt/U)_c$ or $\mu/U -(\mu/U)_c$. Therefore, $g$ for different system sizes should cross at the critical point. As a working hypothesis, we assume that the dynamical exponent $z_c$ equals 1. In Fig. \ref{fig:SStoSF} (a), we show the $\mu/U$ dependence of $g$ at $zt/U=0.24$ and $\beta t=0.5L$. As can be seen in the figure, $g$ actually crosses at a point for different system sizes. From the crossing point, we estimated the quantum critical point as $(\mu/U)_c=0.08455(5)$ for $(zt/U, zV/U)=(0.24, 1)$. 

	To check the consistency of our analysis and clarify the universality class, we analyzed scaling behaviors of $S(\pi, \pi)$ as well as $g$. The scaling form for $S(\pi, \pi)$ is given by $S(\pi, \pi)L^{2\beta_{c}/\nu}=f(\delta L^{1/\nu}, \beta/L^{z_c})$, where $\beta_{c}$ is the critical exponent of the order parameter. Since the effective dimension is $d+z_c=2+1=3$ and the broken symmetry is $Z_2$ symmetry, the quantum phase transition is expected to belong to the 3D Ising universality class. Thus, using the critical exponents $\nu=0.63001$ and $2\beta_{c}/\nu=1.03627$ of the 3D Ising universality class\cite{hasenbusch2010}, we plot $g$ and $S(\pi, \pi)L^{2\beta_{c}/\nu}$ as functions of $\delta L^{1/\nu}$ with fixed $\beta/L$ in Fig. \ref{fig:SStoSF} (b) and (c) respectively. As can be seen in the figure, the data collapses for large system sizes agrees with the expected scaling behavior.
	
\begin{figure}[h]
		\includegraphics[width=7cm]{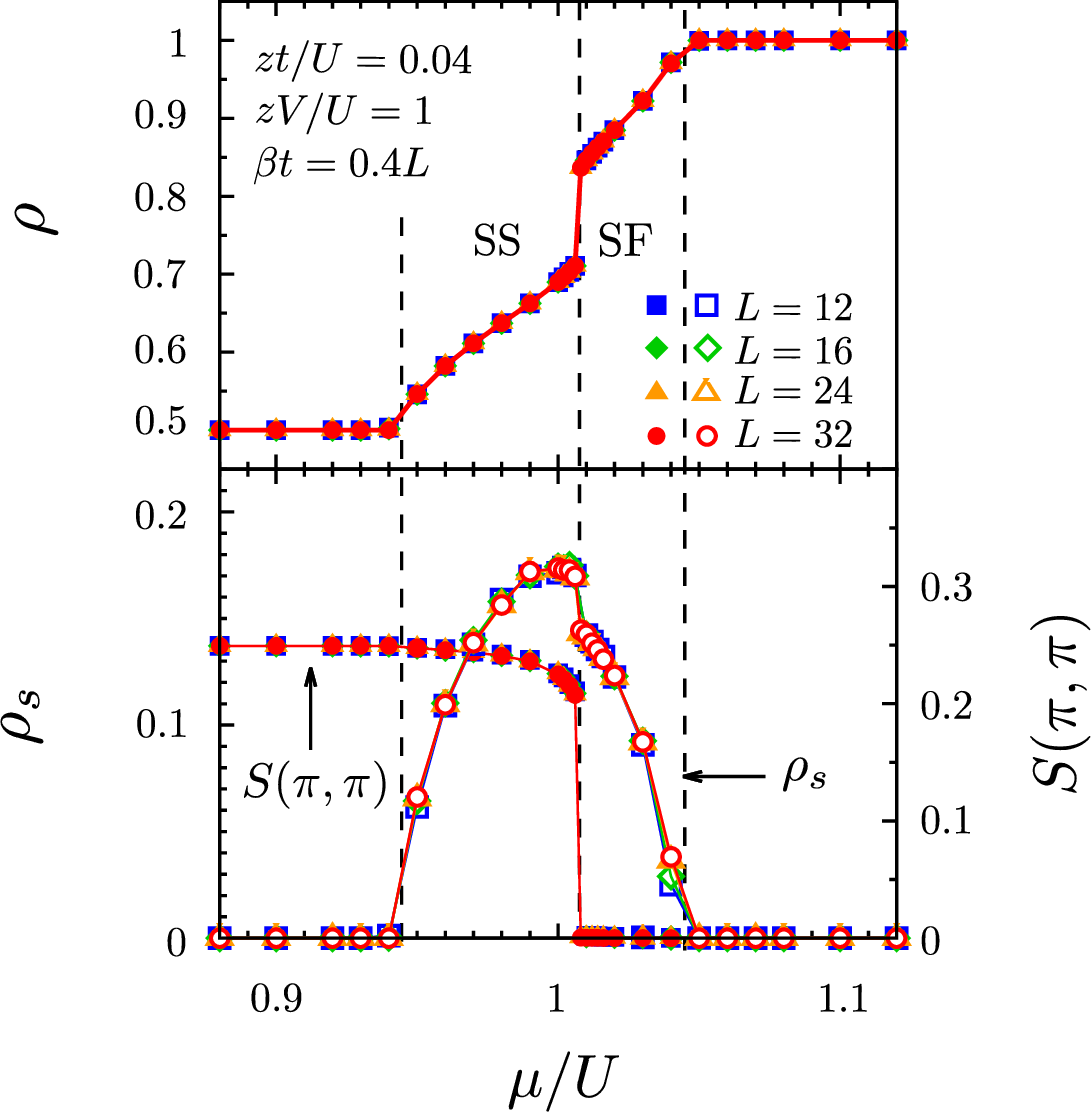}
		\caption{\label{fig:density_jump} (Color online) Finite jump in the particle density at the SS-SF boundary for a small hopping parameter $zt/U=0.04$. Dashed vertical lines are used to separate different phases. In the classical limit $zt/U=0$, the particle density changes discontinuously from 1/2 to 1 at $(\mu/U)_c=1$.}
	\end{figure}

	Exceptional determination of the SS-SF boundaries was made for small hopping parameters $zt/U \alt 0.08$ at $zV/U=1$, because we observed clear finite jumps in the particle density. Fig. \ref{fig:density_jump} shows a jump at the SS-SF boundary, indicating a first-order transition. Similar discontinuities have been also found in the previous quantum Monte Carlo study\cite{sengupta2005}. In this region, we determined the boundary from the position of the jump at low temperatures. The discontinuities of the SS-SF boundaries seem to be connected to ones of the CB-MI boundaries in the classical limit $zt/U=0$ where the particle density changes discontinuously from 1/2 to 1, 1 to 3/2,... at the critical points $(\mu/U)_c=1, 2,...$ respectively. In fact, when the hopping parameter is smaller, the SS-SF transition points approach the classical critical points, as seen in Fig. \ref{fig:phasediag1} (a), and, we found that the finite jump becomes larger. 

\section{\label{sec5}COMMENSURATE SUPERSOLID PHASE}
	Most supersolids are realized by adding/removing particles to/from a commensurate insulating solid. When doped defects delocalize against phase separations and give rise to superfulidity on solid, a supersolid state appears. In contrast to this superolid, the situation of the supersolid at commensurate filling factors is different, because any dopants are absent. In this section, by obtaining simulation results in the canonical ensemble, we investigate supersolid exactly at the commensurate filling factor $\rho=1/2$. To obtain results in the canonical ensemble with the grand-canonical method, we performed the following procedures. We first estimated the chemical potential that corresponds to the desired particle density with high accuracy. Then, we performed simulations at the obtained chemical potential and used only samples whose particle density is exactly equal to the desired one. With this method, in Sec. \ref{subsec5-1}, we obtain direct evidence for supersolid at half-filling, excluding the possibility of phase separations. In the following Sec. \ref{subsec5-2}, we present the ground-state phase diagram at half-filling. The obtained phase diagram shows that the supersolid phase can be found more clearly as the nearest-neighbor repulsion $zV/U$ increases, as the suggestion by the work based on Gutzwiller approximation\cite{kimura2011}.

\subsection{\label{subsec5-1}Supersolid at half-filling}
\begin{figure}[h]
		\includegraphics[width=8cm]{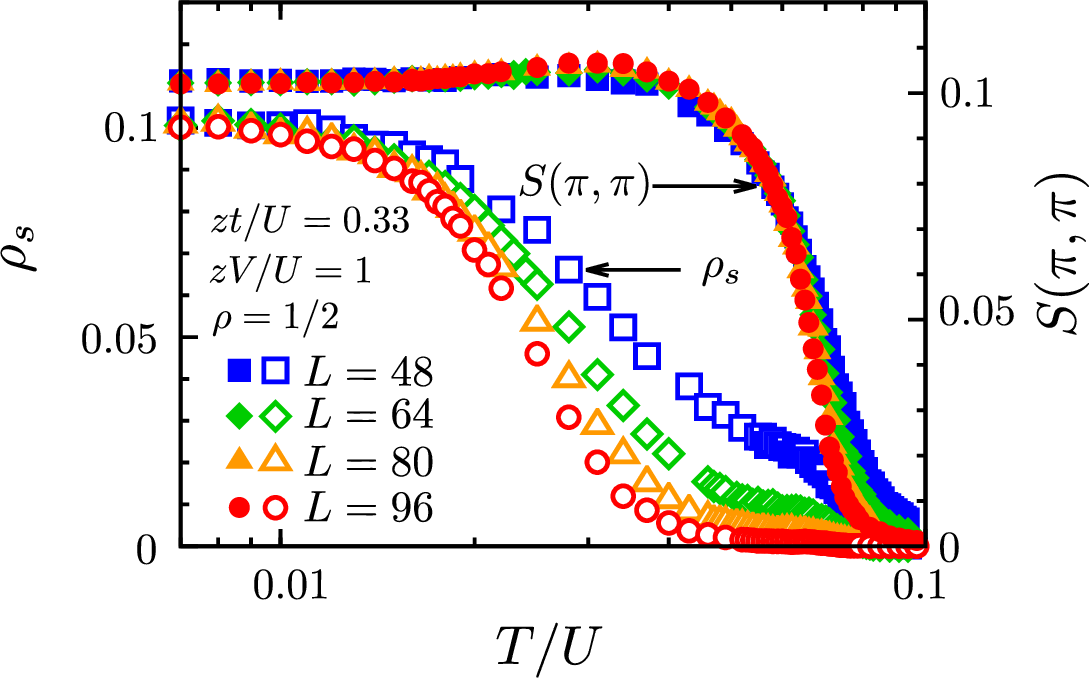}
		\caption{\label{fig:ss_finiteT} (Color online) Finite-temperature dependence of $\rho_s$ and $S(\pi, \pi)$ exactly at half-filling.}
	\end{figure}

\begin{figure}[h]
		\includegraphics[width=4cm]{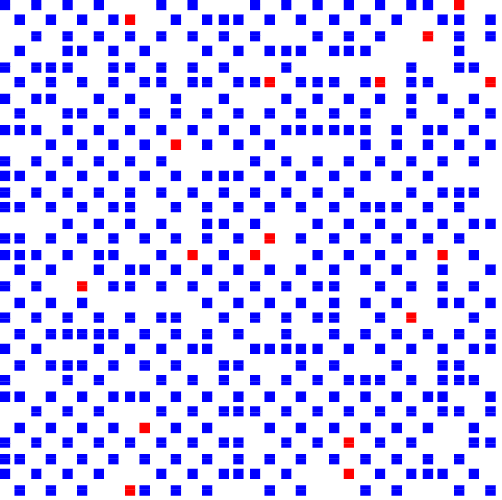}
		\caption{\label{fig:snapshot} (Color online) Snapshot of supersolid at half-filling. This shows a typical configuration in a real space at some particular imaginary time. The parameters are chosen at $(L, zt/U, zV/U, T/U)=(32, 0.33, 1, 0.008)$. Each site are donated as a square. Empty, blue, and red squares indicate empty sites, singly-occupied sites, and doubly-occupied sites respectively.}
	\end{figure}

	In this subsection, we explicitly show the presence of supersolid at half-filling. In Fig. \ref{fig:ss_finiteT}, we plot $\rho_s$ and $S(\pi, \pi)$ as functions of the temperature at half-filling. At low temperatures, both $\rho_s$ and $S(\pi, \pi)$ have finite values, indicating a supersolid state. To exclude the possibility of phase separations, we show a snapshot of the typical configuration in Fig. \ref{fig:snapshot}. In our snapshots, we do not find any macroscopic phase separations. Instead, we can see that the checkerboard solid has microscopic defects (intersitials or vacancies), suggesting the superfluidity is caused by delocalizing defects in the same way as the ordinary supersolids. However, the origin of defects seems to be different from the ordinary one, because it is realized without any change from the commensurate filling factor. Since the CB-to-SS transition at half-filling corresponds to the special transition at the tip of the CB lobe in the grand-canonical phase diagram, it is driven not by adding or subtracting a particle, but by delocalizing quantum fluctuation. Therefore, it is reasonable to interpret the origin of defects as unbound interstitial-vacancy pairs due to the delocalizing quantum fluctuation\cite{prokofev2005}.

\begin{figure}[h]
		\includegraphics[width=9cm]{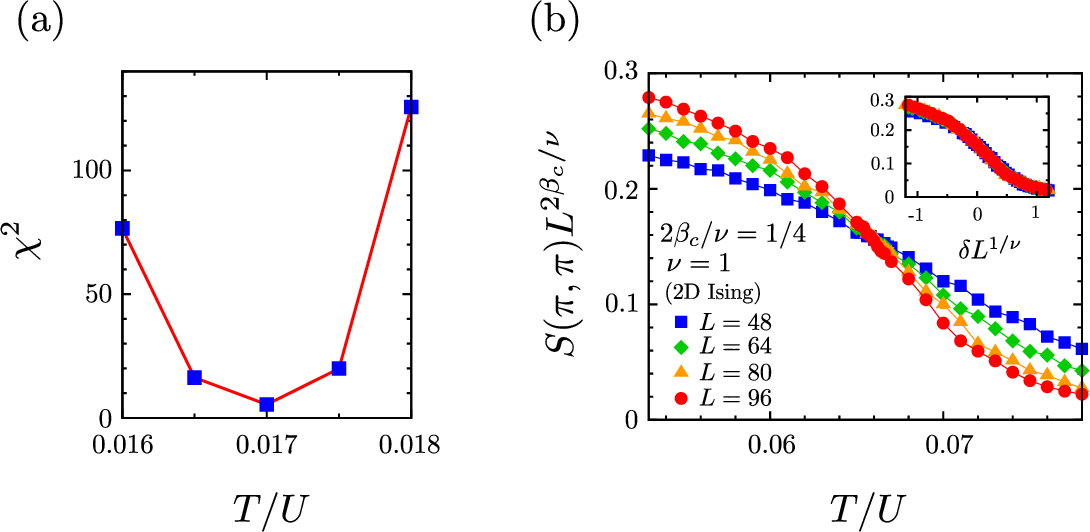}
		\caption{\label{fig:fss_finiteT} (Color online) Determinations of the two critical temperatures in the supersolid state. (a) Values of $\chi^2$ (solid squares) for each temperature. At the critical temperature, the value of $\chi^2$ is expected to be minimized. (b) Intersection of the structure factor $S(\pi, \pi) L^{2\beta_{c}/\nu}$ for different system sizes. The position of the intersection corresponds to the critical temperature of the checkerboard-solid transition. In the inset, we present the data collapse of the scaling plots.}
	\end{figure}

	Melting of the supersolid occurs through two successive finite-temperature transitions, namely superfluid transition and solid transition. Each critical temperature can be determined as follows.
We first consider the superfluid transition. In Fig. \ref{fig:ss_finiteT}, we can observe the strong system size dependence of $\rho_s$ above the superfluid region, which is characteristic of the Kosterlitz-Thouless(KT) transition\cite{kosterlitz1973, kosterlitz1974}. To determine the critical temperature of the KT transition, we make the $\chi^2$ fit to the critical form for the squared winding number\cite{weber1988, harada1998}. Specifically, the squared winding number follows the scaling from of $(\pi/4)\langle \mbox{\boldmath $W$}^2\rangle=1+[2\ln(L/L_0)]^{-1}$ at the critical point. Here, $L_0$ is the only free parameter. For each temperature, we make the $\chi^2$ fit to the critical form and measure the $\chi^2$. Finally, we can obtain a critical temperature as the temperature that minimizes the value of $\chi^2$. The result is shown in Fig. \ref{fig:fss_finiteT} (a). From this analysis, we estimated the critical temperature of the KT transition as $(T/U)_{c}=0.0170(5)$.

	Next, we determined the critical temperature of the checkerboard-solid transition from the structure factor. For finite-temperature phase transitions, the scaling form is given by $S(\pi, \pi) L^{2\beta_{c}/\nu} = f(\delta L^{1/\nu})$, where $\delta=(T/U)-(T/U)_c$. Since the transition is related to the $Z_2$ symmetry breaking, we expect that the critical exponents $2\beta_{c}/\nu$ and $\nu$ equal 1/4 and 1 respectively for the 2D Ising universality class. When this is the case, $S(\pi, \pi) L^{2\beta_{c}/\nu}$ for different system sizes should cross at a critical temperature. Fig. \ref{fig:fss_finiteT} (b) shows the result. In the inset, to check the consistency on the critical exponents, we present the result of the scaling plots that shows the excellent data collapse. Therefore, we obtained the critical temperature of the checkerboard-solid transition as $(T/U)_{c}=0.066(1)$ from the intersecton of $S(\pi, \pi) L^{2\beta_{c}/\nu}$.

	\subsection{\label{subsec5-2}Ground-state phase diagram at half-filling}
	
		In the previous quantum Monte Carlo study\cite{sengupta2005}, supersolid phase has not been found at half-filling for $zt/U=0.2$\cite{sengupta2005}. According to the results from the Gutzwiller approximation, this might be because the hopping parameter is not sufficiently large for supersolid to be found clearly at half-filling\cite{kimura2011}. In this subsection, to confirm this suggestion, we clarify the parameter dependence of the supersolid region at half-filling.

\begin{figure}[h]
		\includegraphics[width=6.5cm]{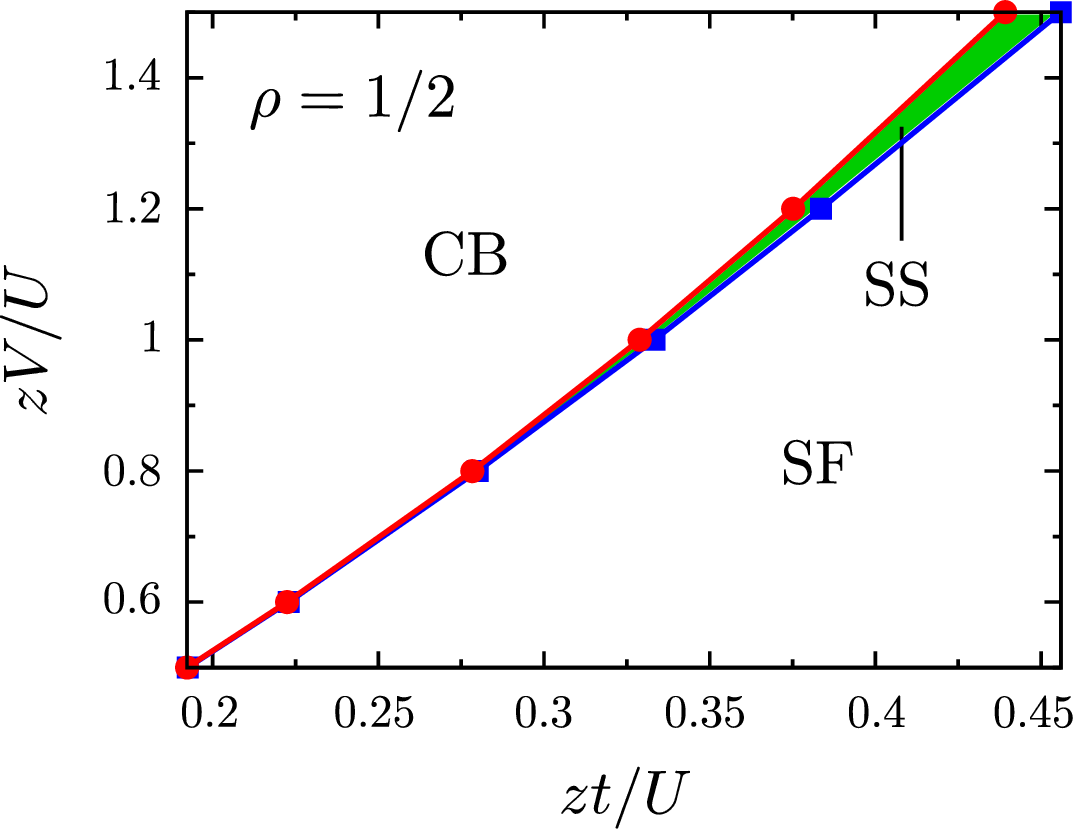}
		\caption{\label{fig:phasediag_canonical} (Color online) Ground-state phase diagram at half-filling. Circles and squares denote critical points which correspond to onsets of checkerboard order and superfluid respectively. The lines are used to guide the eyes. The green region between the two lines represents the supersolid (SS) phase.}
	\end{figure}

\begin{figure}[h]
		\includegraphics[width=7cm]{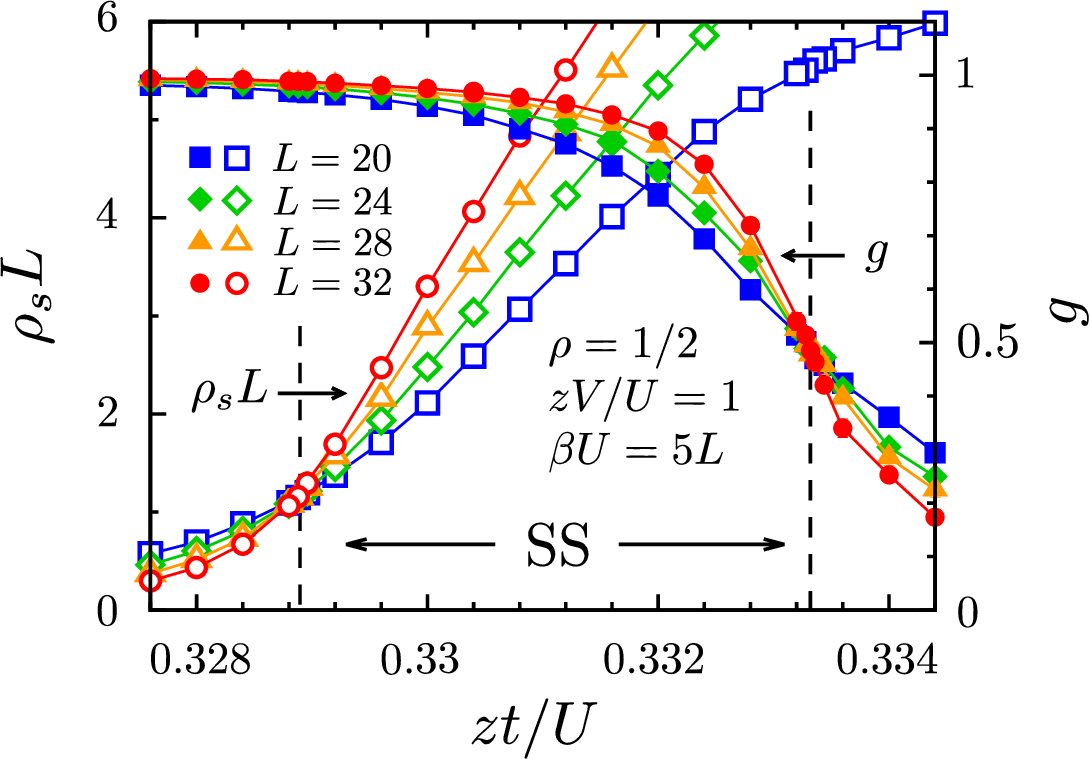}
		\caption{\label{fig:boundary_canonical} (Color online) Estimation of quantum critical points from intersection of $\rho_s L$ or $g$ for different system sizes. Dashed vertical lines are placed at the estimated critical points for the CB-SS transition (left) and SS-SF transition (right).}
	\end{figure}

		In Fig. \ref{fig:phasediag_canonical}, we present the ground-state phase diagram at half-filling in the $zt/U$-$zV/U$ plane. The phase boundaries are determined from the position of an intersection of $g$ or $\rho_s L$ for different system sizes with the assumption that $z_c$ equals 1. Fig. \ref{fig:boundary_canonical} shows a result at $zV/U=1$. In the figure, we obtained the quantum critical points for the CB-SS transition and the SS-SF transition as $(zt/U)_c=0.32888(8)$ and 0.33332(8) respectively. Note that the critical point for the CB-SS transition at $\rho=1/2$ agrees with that obtained from the grand-canonical ensemble(Sec. \ref{subsec4-2}). In our phase diagram, the supersolid region is much smaller than that obtained by the Gutzwiller approximation\cite{kimura2011}. However, qualitative behaviors of the phase boundaries agree with the Gutzwiller results. As the nearest-neighbor repulsion $zV/U$ increases, the CB phase expands up to larger hopping parameters $zt/U$. The SS phase also extends for large nearest-neighbor repulsions and hopping parameters. In contrast, for the small hopping parameters including $zt/U=0.2$, the two phase boundaries are very close to each other. Thus, we conclude that the reason why the SS phase was not found at half-filling in the previous quantum Monte Carlo result\cite{sengupta2005} is that the hopping parameter used was not enough large for the SS phase to be observed clearly, as the author of Ref.\cite{kimura2011} predicted.
	
\section{\label{sec6}SUMMARY}
	In conclusion, we have investigated the ground-state phase diagrams of the 2D extended Bose-Hubbard model by performing unbiased quantum Monte Carlo simulations. Especially, we find that the ground-state phase diagrams by Gutzwiller mean-field approximation are qualitatively correct and the supersolid below and at half-filling are stable as well as the 3D system. For the strong nearest-neighbor repulsion, we have also confirmed that the supersolid phase exits up to large hopping parameters. Although the 2D result qualitatively agrees with the 3D or Gutzwiller mean-field results, the supersolid regions shrinks in the lower dimensions due to the quantum fluctuation.

\section*{ACKNOWLEDGMENTS}
	This work was financially supported by the Global COE Program ``the Physical Science Frontier", the Grant-in-Aid for JSPS Fellows (Grant No. 249904), the Grant-in-Aid for Scientific Research (B) (22340111), the Computational Materials Science Initiative (CMSI), Japan. The simulations were performed on computers at the Supercomputer Center, Institute for Solid State Physics, University of Tokyo.


\begin{thebibliography}{51}
\expandafter\ifx\csname natexlab\endcsname\relax\def\natexlab#1{#1}\fi
\expandafter\ifx\csname bibnamefont\endcsname\relax
  \def\bibnamefont#1{#1}\fi
\expandafter\ifx\csname bibfnamefont\endcsname\relax
  \def\bibfnamefont#1{#1}\fi
\expandafter\ifx\csname citenamefont\endcsname\relax
  \def\citenamefont#1{#1}\fi
\expandafter\ifx\csname url\endcsname\relax
  \def\url#1{\texttt{#1}}\fi
\expandafter\ifx\csname urlprefix\endcsname\relax\def\urlprefix{URL }\fi
\providecommand{\bibinfo}[2]{#2}
\providecommand{\eprint}[2][]{\url{#2}}

\bibitem[{\citenamefont{Andreev and Lifshitz}(1969)}]{andreev1969}
\bibinfo{author}{\bibfnamefont{A.~F.} \bibnamefont{Andreev}} \bibnamefont{and}
  \bibinfo{author}{\bibfnamefont{I.~M.} \bibnamefont{Lifshitz}},
  \bibinfo{journal}{Sov. Phys. JETP} \textbf{\bibinfo{volume}{29}},
  \bibinfo{pages}{1107} (\bibinfo{year}{1969}).

\bibitem[{\citenamefont{Chester}(1970)}]{chester1970}
\bibinfo{author}{\bibfnamefont{G.~V.} \bibnamefont{Chester}},
  \bibinfo{journal}{Phys. Rev. A} \textbf{\bibinfo{volume}{2}},
  \bibinfo{pages}{256} (\bibinfo{year}{1970}).

\bibitem[{\citenamefont{Kim and Chan}(2004{\natexlab{a}})}]{kim2004-1}
\bibinfo{author}{\bibfnamefont{E.}~\bibnamefont{Kim}} \bibnamefont{and}
  \bibinfo{author}{\bibfnamefont{M.~H.~W.} \bibnamefont{Chan}},
  \bibinfo{journal}{Nature (London)} \textbf{\bibinfo{volume}{427}},
  \bibinfo{pages}{225} (\bibinfo{year}{2004}{\natexlab{a}}).

\bibitem[{\citenamefont{Kim and Chan}(2004{\natexlab{b}})}]{kim2004-2}
\bibinfo{author}{\bibfnamefont{E.}~\bibnamefont{Kim}} \bibnamefont{and}
  \bibinfo{author}{\bibfnamefont{M.~H.~W.} \bibnamefont{Chan}},
  \bibinfo{journal}{Science} \textbf{\bibinfo{volume}{305}},
  \bibinfo{pages}{1941} (\bibinfo{year}{2004}{\natexlab{b}}).

\bibitem[{\citenamefont{Prokof'ev and Svistunov}(2005)}]{prokofev2005}
\bibinfo{author}{\bibfnamefont{N.}~\bibnamefont{Prokof'ev}} \bibnamefont{and}
  \bibinfo{author}{\bibfnamefont{B.}~\bibnamefont{Svistunov}},
  \bibinfo{journal}{Phys. Rev. Lett.} \textbf{\bibinfo{volume}{94}},
  \bibinfo{pages}{155302} (\bibinfo{year}{2005}).

\bibitem[{\citenamefont{Boninsegni et~al.}(2006)\citenamefont{Boninsegni,
  Kuklov, Pollet, Prokof'ev, Svistunov, and Troyer}}]{boninsegni2006}
\bibinfo{author}{\bibfnamefont{M.}~\bibnamefont{Boninsegni}},
  \bibinfo{author}{\bibfnamefont{A.~B.} \bibnamefont{Kuklov}},
  \bibinfo{author}{\bibfnamefont{L.}~\bibnamefont{Pollet}},
  \bibinfo{author}{\bibfnamefont{N.~V.} \bibnamefont{Prokof'ev}},
  \bibinfo{author}{\bibfnamefont{B.}~\bibnamefont{Svistunov}},
  \bibnamefont{and} \bibinfo{author}{\bibfnamefont{M.}~\bibnamefont{Troyer}},
  \bibinfo{journal}{Phys. Rev. Lett.} \textbf{\bibinfo{volume}{97}},
  \bibinfo{pages}{080401} (\bibinfo{year}{2006}).

\bibitem[{\citenamefont{Rittner and Reppy}(2006)}]{rittner2006}
\bibinfo{author}{\bibfnamefont{A.~S.~C.} \bibnamefont{Rittner}}
  \bibnamefont{and} \bibinfo{author}{\bibfnamefont{J.~D.} \bibnamefont{Reppy}},
  \bibinfo{journal}{Phys. Rev. Lett.} \textbf{\bibinfo{volume}{97}},
  \bibinfo{pages}{165301} (\bibinfo{year}{2006}).

\bibitem[{\citenamefont{Sasaki et~al.}(2006)\citenamefont{Sasaki, Ishiguro,
  Caupin, Maris, and Balibar}}]{sasaki2006}
\bibinfo{author}{\bibfnamefont{S.}~\bibnamefont{Sasaki}},
  \bibinfo{author}{\bibfnamefont{R.}~\bibnamefont{Ishiguro}},
  \bibinfo{author}{\bibfnamefont{F.}~\bibnamefont{Caupin}},
  \bibinfo{author}{\bibfnamefont{H.~J.} \bibnamefont{Maris}}, \bibnamefont{and}
  \bibinfo{author}{\bibfnamefont{S.}~\bibnamefont{Balibar}},
  \bibinfo{journal}{Science} \textbf{\bibinfo{volume}{313}},
  \bibinfo{pages}{1098} (\bibinfo{year}{2006}).

\bibitem[{\citenamefont{Pollet et~al.}(2007)\citenamefont{Pollet, Boninsegni,
  Kuklov, Prokof'ev, Svisunov, and Troyer}}]{pollet2007}
\bibinfo{author}{\bibfnamefont{L.}~\bibnamefont{Pollet}},
  \bibinfo{author}{\bibfnamefont{M.}~\bibnamefont{Boninsegni}},
  \bibinfo{author}{\bibfnamefont{A.~B.} \bibnamefont{Kuklov}},
  \bibinfo{author}{\bibfnamefont{N.~V.} \bibnamefont{Prokof'ev}},
  \bibinfo{author}{\bibfnamefont{B.~V.} \bibnamefont{Svisunov}},
  \bibnamefont{and} \bibinfo{author}{\bibfnamefont{M.}~\bibnamefont{Troyer}},
  \bibinfo{journal}{Phys. Rev. Lett.} \textbf{\bibinfo{volume}{98}},
  \bibinfo{pages}{135301} (\bibinfo{year}{2007}).

\bibitem[{\citenamefont{Boninsegni et~al.}(2007)\citenamefont{Boninsegni,
  Kuklov, Pollet, Prokof'ev, Svistunov, and Troyer}}]{boninsegni2007}
\bibinfo{author}{\bibfnamefont{M.}~\bibnamefont{Boninsegni}},
  \bibinfo{author}{\bibfnamefont{A.~B.} \bibnamefont{Kuklov}},
  \bibinfo{author}{\bibfnamefont{L.}~\bibnamefont{Pollet}},
  \bibinfo{author}{\bibfnamefont{N.~V.} \bibnamefont{Prokof'ev}},
  \bibinfo{author}{\bibfnamefont{B.~V.} \bibnamefont{Svistunov}},
  \bibnamefont{and} \bibinfo{author}{\bibfnamefont{M.}~\bibnamefont{Troyer}},
  \bibinfo{journal}{Phys. Rev. Lett.} \textbf{\bibinfo{volume}{99}},
  \bibinfo{pages}{035301} (\bibinfo{year}{2007}).

\bibitem[{\citenamefont{Greiner et~al.}(2002)\citenamefont{Greiner, Mandel,
  Esslinger, H$\ddot{{\rm a}}$nsch, and Bloch}}]{greiner2002}
\bibinfo{author}{\bibfnamefont{M.}~\bibnamefont{Greiner}},
  \bibinfo{author}{\bibfnamefont{O.}~\bibnamefont{Mandel}},
  \bibinfo{author}{\bibfnamefont{T.}~\bibnamefont{Esslinger}},
  \bibinfo{author}{\bibfnamefont{T.~W.} \bibnamefont{H$\ddot{{\rm a}}$nsch}},
  \bibnamefont{and} \bibinfo{author}{\bibfnamefont{I.}~\bibnamefont{Bloch}},
  \bibinfo{journal}{Nature} \textbf{\bibinfo{volume}{415}}, \bibinfo{pages}{39}
  (\bibinfo{year}{2002}).

\bibitem[{\citenamefont{Sage et~al.}(2005)\citenamefont{Sage, Sainis, Bergeman,
  and DeMille}}]{sage2005}
\bibinfo{author}{\bibfnamefont{J.~M.} \bibnamefont{Sage}},
  \bibinfo{author}{\bibfnamefont{S.}~\bibnamefont{Sainis}},
  \bibinfo{author}{\bibfnamefont{T.}~\bibnamefont{Bergeman}}, \bibnamefont{and}
  \bibinfo{author}{\bibfnamefont{D.}~\bibnamefont{DeMille}},
  \bibinfo{journal}{Phys. Rev. Lett.} \textbf{\bibinfo{volume}{94}},
  \bibinfo{pages}{203001} (\bibinfo{year}{2005}).

\bibitem[{\citenamefont{Ni et~al.}(2008)\citenamefont{Ni, Ospelkaus, Miranda,
  Peer, Neyenhuis, Zirbel, S.Kotochigova, Julienne, Jin, and Ye}}]{ni2008}
\bibinfo{author}{\bibfnamefont{K.-K.} \bibnamefont{Ni}},
  \bibinfo{author}{\bibfnamefont{S.}~\bibnamefont{Ospelkaus}},
  \bibinfo{author}{\bibfnamefont{M.~H.~G.} \bibnamefont{Miranda}},
  \bibinfo{author}{\bibfnamefont{A.}~\bibnamefont{Peer}},
  \bibinfo{author}{\bibfnamefont{B.}~\bibnamefont{Neyenhuis}},
  \bibinfo{author}{\bibfnamefont{J.~J.} \bibnamefont{Zirbel}},
  \bibinfo{author}{\bibnamefont{S.Kotochigova}},
  \bibinfo{author}{\bibfnamefont{P.~S.} \bibnamefont{Julienne}},
  \bibinfo{author}{\bibfnamefont{D.~S.} \bibnamefont{Jin}}, \bibnamefont{and}
  \bibinfo{author}{\bibfnamefont{J.}~\bibnamefont{Ye}},
  \bibinfo{journal}{Science} \textbf{\bibinfo{volume}{322}},
  \bibinfo{pages}{231} (\bibinfo{year}{2008}).

\bibitem[{\citenamefont{Ospelkaus et~al.}(2008)\citenamefont{Ospelkaus,
  P${\acute{\rm e}}$er, Zirbel, Neyenhuis, Kotochigova, Julienne, Ye, and
  Jin}}]{ospelkaus2008}
\bibinfo{author}{\bibfnamefont{S.}~\bibnamefont{Ospelkaus}},
  \bibinfo{author}{\bibfnamefont{A.}~\bibnamefont{P${\acute{\rm e}}$er}},
  \bibinfo{author}{\bibfnamefont{J.~J.} \bibnamefont{Zirbel}},
  \bibinfo{author}{\bibfnamefont{B.}~\bibnamefont{Neyenhuis}},
  \bibinfo{author}{\bibfnamefont{S.}~\bibnamefont{Kotochigova}},
  \bibinfo{author}{\bibfnamefont{P.~S.} \bibnamefont{Julienne}},
  \bibinfo{author}{\bibfnamefont{J.}~\bibnamefont{Ye}}, \bibnamefont{and}
  \bibinfo{author}{\bibfnamefont{D.~S.} \bibnamefont{Jin}},
  \bibinfo{journal}{Nature Phys.} \textbf{\bibinfo{volume}{4}},
  \bibinfo{pages}{622} (\bibinfo{year}{2008}).

\bibitem[{\citenamefont{Jaksch et~al.}(1998)\citenamefont{Jaksch, Bruder,
  Cirac, Gardiner, and Zoller}}]{jaksch1998}
\bibinfo{author}{\bibfnamefont{D.}~\bibnamefont{Jaksch}},
  \bibinfo{author}{\bibfnamefont{C.}~\bibnamefont{Bruder}},
  \bibinfo{author}{\bibfnamefont{J.~I.} \bibnamefont{Cirac}},
  \bibinfo{author}{\bibfnamefont{C.~W.} \bibnamefont{Gardiner}},
  \bibnamefont{and} \bibinfo{author}{\bibfnamefont{P.}~\bibnamefont{Zoller}},
  \bibinfo{journal}{Phys. Rev. Lett.} \textbf{\bibinfo{volume}{81}},
  \bibinfo{pages}{3108} (\bibinfo{year}{1998}).

\bibitem[{\citenamefont{van Otterlo et~al.}(2005)\citenamefont{van Otterlo,
  Wagenblast, Blatin, Bruder, Fazio, and Sch$\ddot{{\rm o}}$n}}]{otterlo1995}
\bibinfo{author}{\bibfnamefont{A.}~\bibnamefont{van Otterlo}},
  \bibinfo{author}{\bibfnamefont{K.~H.} \bibnamefont{Wagenblast}},
  \bibinfo{author}{\bibfnamefont{R.}~\bibnamefont{Blatin}},
  \bibinfo{author}{\bibfnamefont{C.}~\bibnamefont{Bruder}},
  \bibinfo{author}{\bibfnamefont{R.}~\bibnamefont{Fazio}}, \bibnamefont{and}
  \bibinfo{author}{\bibfnamefont{G.}~\bibnamefont{Sch$\ddot{{\rm o}}$n}},
  \bibinfo{journal}{Phys. Rev. B} \textbf{\bibinfo{volume}{52}},
  \bibinfo{pages}{16176} (\bibinfo{year}{2005}).

\bibitem[{\citenamefont{Batrouni and Scalettar}(2000)}]{batrouni2000}
\bibinfo{author}{\bibfnamefont{G.~G.} \bibnamefont{Batrouni}} \bibnamefont{and}
  \bibinfo{author}{\bibfnamefont{R.~T.} \bibnamefont{Scalettar}},
  \bibinfo{journal}{Phys. Rev. Lett.} \textbf{\bibinfo{volume}{84}},
  \bibinfo{pages}{1599} (\bibinfo{year}{2000}).

\bibitem[{\citenamefont{G$\acute{{\rm o}}$ral
  et~al.}(2002)\citenamefont{G$\acute{{\rm o}}$ral, Santos, and
  Lewenstein}}]{goral2002}
\bibinfo{author}{\bibfnamefont{K.}~\bibnamefont{G$\acute{{\rm o}}$ral}},
  \bibinfo{author}{\bibfnamefont{L.}~\bibnamefont{Santos}}, \bibnamefont{and}
  \bibinfo{author}{\bibfnamefont{M.}~\bibnamefont{Lewenstein}},
  \bibinfo{journal}{Phys. Rev. Lett.} \textbf{\bibinfo{volume}{88}},
  \bibinfo{pages}{170406} (\bibinfo{year}{2002}).

\bibitem[{\citenamefont{Wessel and Troyer}(2005)}]{wessel2005}
\bibinfo{author}{\bibfnamefont{S.}~\bibnamefont{Wessel}} \bibnamefont{and}
  \bibinfo{author}{\bibfnamefont{M.}~\bibnamefont{Troyer}},
  \bibinfo{journal}{Phys. Rev. Lett.} \textbf{\bibinfo{volume}{95}},
  \bibinfo{pages}{127205} (\bibinfo{year}{2005}).

\bibitem[{\citenamefont{Boninsegni and Prokof'ev}(2005)}]{boninsegni2005}
\bibinfo{author}{\bibfnamefont{M.}~\bibnamefont{Boninsegni}} \bibnamefont{and}
  \bibinfo{author}{\bibfnamefont{N.}~\bibnamefont{Prokof'ev}},
  \bibinfo{journal}{Phys. Rev. Lett.} \textbf{\bibinfo{volume}{95}},
  \bibinfo{pages}{237204} (\bibinfo{year}{2005}).

\bibitem[{\citenamefont{Sengupta et~al.}(2005)\citenamefont{Sengupta, Pryadko,
  Alet, Troyer, and Schmid}}]{sengupta2005}
\bibinfo{author}{\bibfnamefont{P.}~\bibnamefont{Sengupta}},
  \bibinfo{author}{\bibfnamefont{L.~P.} \bibnamefont{Pryadko}},
  \bibinfo{author}{\bibfnamefont{F.}~\bibnamefont{Alet}},
  \bibinfo{author}{\bibfnamefont{M.}~\bibnamefont{Troyer}}, \bibnamefont{and}
  \bibinfo{author}{\bibfnamefont{G.}~\bibnamefont{Schmid}},
  \bibinfo{journal}{Phys. Rev. Lett.} \textbf{\bibinfo{volume}{94}},
  \bibinfo{pages}{207202} (\bibinfo{year}{2005}).

\bibitem[{\citenamefont{Batrouni et~al.}(2006)\citenamefont{Batrouni,
  H$\acute{\rm e}$bert, and Scalettar}}]{batrouni2006}
\bibinfo{author}{\bibfnamefont{G.~G.} \bibnamefont{Batrouni}},
  \bibinfo{author}{\bibfnamefont{F.}~\bibnamefont{H$\acute{\rm e}$bert}},
  \bibnamefont{and} \bibinfo{author}{\bibfnamefont{R.~T.}
  \bibnamefont{Scalettar}}, \bibinfo{journal}{Phys. Rev. Lett.}
  \textbf{\bibinfo{volume}{97}}, \bibinfo{pages}{087209}
  (\bibinfo{year}{2006}).

\bibitem[{\citenamefont{Yi et~al.}(2007)\citenamefont{Yi, Li, and
  Sun}}]{yi2007}
\bibinfo{author}{\bibfnamefont{S.}~\bibnamefont{Yi}},
  \bibinfo{author}{\bibfnamefont{T.}~\bibnamefont{Li}}, \bibnamefont{and}
  \bibinfo{author}{\bibfnamefont{C.~P.} \bibnamefont{Sun}},
  \bibinfo{journal}{Phys. Rev. Lett.} \textbf{\bibinfo{volume}{98}},
  \bibinfo{pages}{260405} (\bibinfo{year}{2007}).

\bibitem[{\citenamefont{Suzuki and Kawashima}(2007)}]{suzuki2007}
\bibinfo{author}{\bibfnamefont{T.}~\bibnamefont{Suzuki}} \bibnamefont{and}
  \bibinfo{author}{\bibfnamefont{N.}~\bibnamefont{Kawashima}},
  \bibinfo{journal}{Phys. Rev. B} \textbf{\bibinfo{volume}{75}},
  \bibinfo{pages}{180502(R)} (\bibinfo{year}{2007}).

\bibitem[{\citenamefont{Dang et~al.}(2008)\citenamefont{Dang, Boninsegni, and
  Pollet}}]{dang2008}
\bibinfo{author}{\bibfnamefont{L.}~\bibnamefont{Dang}},
  \bibinfo{author}{\bibfnamefont{M.}~\bibnamefont{Boninsegni}},
  \bibnamefont{and} \bibinfo{author}{\bibfnamefont{L.}~\bibnamefont{Pollet}},
  \bibinfo{journal}{Phys. Rev. B} \textbf{\bibinfo{volume}{78}},
  \bibinfo{pages}{132512} (\bibinfo{year}{2008}).

\bibitem[{\citenamefont{Yamamoto et~al.}(2009)\citenamefont{Yamamoto, Todo, and
  Miyashita}}]{yamamoto2009}
\bibinfo{author}{\bibfnamefont{K.}~\bibnamefont{Yamamoto}},
  \bibinfo{author}{\bibfnamefont{S.}~\bibnamefont{Todo}}, \bibnamefont{and}
  \bibinfo{author}{\bibfnamefont{S.}~\bibnamefont{Miyashita}},
  \bibinfo{journal}{Phys. Rev. B} \textbf{\bibinfo{volume}{79}},
  \bibinfo{pages}{094503} (\bibinfo{year}{2009}).

\bibitem[{\citenamefont{Danshita and de~Melo}(2009)}]{danshita2009}
\bibinfo{author}{\bibfnamefont{I.}~\bibnamefont{Danshita}} \bibnamefont{and}
  \bibinfo{author}{\bibfnamefont{C.~A. R.} \bibnamefont{S$\acute{\rm a}$~de~Melo}},
  \bibinfo{journal}{Phys. Rev. Lett.} \textbf{\bibinfo{volume}{103}},
  \bibinfo{pages}{225301} (\bibinfo{year}{2009}).

\bibitem[{\citenamefont{Pollet et~al.}(2010)\citenamefont{Pollet, Picon,
  B$\ddot{{\rm u}}$chler, and Troyer}}]{pollet2010}
\bibinfo{author}{\bibfnamefont{L.}~\bibnamefont{Pollet}},
  \bibinfo{author}{\bibfnamefont{J.~D.} \bibnamefont{Picon}},
  \bibinfo{author}{\bibfnamefont{H.~P.} \bibnamefont{B$\ddot{{\rm u}}$chler}},
  \bibnamefont{and} \bibinfo{author}{\bibfnamefont{M.}~\bibnamefont{Troyer}},
  \bibinfo{journal}{Phys. Rev. Lett.} \textbf{\bibinfo{volume}{104}},
  \bibinfo{pages}{125302} (\bibinfo{year}{2010}).

\bibitem[{\citenamefont{Capogrosso-Sansone
  et~al.}(2010)\citenamefont{Capogrosso-Sansone, Trefzger, Lewenstein, Zoller,
  and Pupillo}}]{capogrosso2010-1}
\bibinfo{author}{\bibfnamefont{B.}~\bibnamefont{Capogrosso-Sansone}},
  \bibinfo{author}{\bibfnamefont{C.}~\bibnamefont{Trefzger}},
  \bibinfo{author}{\bibfnamefont{M.}~\bibnamefont{Lewenstein}},
  \bibinfo{author}{\bibfnamefont{P.}~\bibnamefont{Zoller}}, \bibnamefont{and}
  \bibinfo{author}{\bibfnamefont{G.}~\bibnamefont{Pupillo}},
  \bibinfo{journal}{Phys. Rev. Lett.} \textbf{\bibinfo{volume}{104}},
  \bibinfo{pages}{125301} (\bibinfo{year}{2010}).

\bibitem[{\citenamefont{Xi et~al.}(2011)\citenamefont{Xi, Ye, Chen, Zhang, and
  Su}}]{xi2011}
\bibinfo{author}{\bibfnamefont{B.}~\bibnamefont{Xi}},
  \bibinfo{author}{\bibfnamefont{F.}~\bibnamefont{Ye}},
  \bibinfo{author}{\bibfnamefont{W.}~\bibnamefont{Chen}},
  \bibinfo{author}{\bibfnamefont{F.}~\bibnamefont{Zhang}}, \bibnamefont{and}
  \bibinfo{author}{\bibfnamefont{G.}~\bibnamefont{Su}}, \bibinfo{journal}{Phys.
  Rev. B} \textbf{\bibinfo{volume}{84}}, \bibinfo{pages}{054512}
  (\bibinfo{year}{2011}).

\bibitem[{\citenamefont{Yamamoto et~al.}(2012)\citenamefont{Yamamoto, Danshita,
  and de~Melo}}]{yamamoto2012}
\bibinfo{author}{\bibfnamefont{D.}~\bibnamefont{Yamamoto}},
  \bibinfo{author}{\bibfnamefont{I.}~\bibnamefont{Danshita}}, \bibnamefont{and}
  \bibinfo{author}{\bibfnamefont{C.~A. R.} \bibnamefont{S$\acute{\rm a}$~de~Melo}},
  \bibinfo{journal}{Phys. Rev. A} \textbf{\bibinfo{volume}{85}},
  \bibinfo{pages}{021601(R)} (\bibinfo{year}{2012}).

\bibitem[{\citenamefont{Ohgoe et~al.}(2012)\citenamefont{Ohgoe, Suzuki, and
  Kawashima}}]{ohgoe2012}
\bibinfo{author}{\bibfnamefont{T.}~\bibnamefont{Ohgoe}},
  \bibinfo{author}{\bibfnamefont{T.}~\bibnamefont{Suzuki}}, \bibnamefont{and}
  \bibinfo{author}{\bibfnamefont{N.}~\bibnamefont{Kawashima}},
  \bibinfo{journal}{Phys. Rev. Lett.} \textbf{\bibinfo{volume}{108}},
  \bibinfo{pages}{185302} (\bibinfo{year}{2012}).

\bibitem[{\citenamefont{Kimura}(2011)}]{kimura2011}
\bibinfo{author}{\bibfnamefont{T.}~\bibnamefont{Kimura}},
  \bibinfo{journal}{Phys. Rev. A} \textbf{\bibinfo{volume}{84}},
  \bibinfo{pages}{063630} (\bibinfo{year}{2011}).

\bibitem[{\citenamefont{Iskin}(2011)}]{iskin2011}
\bibinfo{author}{\bibfnamefont{M.}~\bibnamefont{Iskin}},
  \bibinfo{journal}{Phys. Rev. A} \textbf{\bibinfo{volume}{83}},
  \bibinfo{pages}{051606(R)} (\bibinfo{year}{2011}).

\bibitem[{\citenamefont{Kovrizhin et~al.}(2005)\citenamefont{Kovrizhin, Pai,
  and Shinha}}]{kovrizhin2005}
\bibinfo{author}{\bibfnamefont{D.~L.} \bibnamefont{Kovrizhin}},
  \bibinfo{author}{\bibfnamefont{G.~V.} \bibnamefont{Pai}}, \bibnamefont{and}
  \bibinfo{author}{\bibfnamefont{S.}~\bibnamefont{Shinha}},
  \bibinfo{journal}{Europhys. Lett.} \textbf{\bibinfo{volume}{72}}
  (\bibinfo{year}{2005}).

\bibitem[{\citenamefont{Griesmaier et~al.}(2005)\citenamefont{Griesmaier,
  Werner, Hensler, Stuhler, and Pfau}}]{griesmaier2005}
\bibinfo{author}{\bibfnamefont{A.}~\bibnamefont{Griesmaier}},
  \bibinfo{author}{\bibfnamefont{J.}~\bibnamefont{Werner}},
  \bibinfo{author}{\bibfnamefont{S.}~\bibnamefont{Hensler}},
  \bibinfo{author}{\bibfnamefont{J.}~\bibnamefont{Stuhler}}, \bibnamefont{and}
  \bibinfo{author}{\bibfnamefont{T.}~\bibnamefont{Pfau}},
  \bibinfo{journal}{Phys. Rev. Lett.} \textbf{\bibinfo{volume}{94}},
  \bibinfo{pages}{160401} (\bibinfo{year}{2005}).

\bibitem[{\citenamefont{Prokof'ev et~al.}(1998)\citenamefont{Prokof'ev,
  Svistunov, and Tupitsyn}}]{prokofev1998}
\bibinfo{author}{\bibfnamefont{N.~V.} \bibnamefont{Prokof'ev}},
  \bibinfo{author}{\bibfnamefont{B.~V.} \bibnamefont{Svistunov}},
  \bibnamefont{and} \bibinfo{author}{\bibfnamefont{I.~S.}
  \bibnamefont{Tupitsyn}}, \bibinfo{journal}{Sov. Phys. JETP}
  \textbf{\bibinfo{volume}{87}}, \bibinfo{pages}{310} (\bibinfo{year}{1998}).

\bibitem[{\citenamefont{Sylju{\aa}sen and Sandvik}(2002)}]{syljuasen2002}
\bibinfo{author}{\bibfnamefont{O.~F.} \bibnamefont{Sylju{\aa}sen}}
  \bibnamefont{and} \bibinfo{author}{\bibfnamefont{A.~W.}
  \bibnamefont{Sandvik}}, \bibinfo{journal}{Phys. Rev. E}
  \textbf{\bibinfo{volume}{66}}, \bibinfo{pages}{046701}
  (\bibinfo{year}{2002}).

\bibitem[{\citenamefont{Kawashima and Harada}(2004)}]{kawashima2004}
\bibinfo{author}{\bibfnamefont{N.}~\bibnamefont{Kawashima}} \bibnamefont{and}
  \bibinfo{author}{\bibfnamefont{K.}~\bibnamefont{Harada}},
  \bibinfo{journal}{J. Phys. Soc. Jpn.} \textbf{\bibinfo{volume}{73}},
  \bibinfo{pages}{1379} (\bibinfo{year}{2004}).

\bibitem[{\citenamefont{Kato and Kawashima}(2009)}]{kato2009}
\bibinfo{author}{\bibfnamefont{Y.}~\bibnamefont{Kato}} \bibnamefont{and}
  \bibinfo{author}{\bibfnamefont{N.}~\bibnamefont{Kawashima}},
  \bibinfo{journal}{Phys. Rev. E} \textbf{\bibinfo{volume}{79}},
  \bibinfo{pages}{021104} (\bibinfo{year}{2009}).

\bibitem[{\citenamefont{Pollock and Ceperley}(1987)}]{pollock1987}
\bibinfo{author}{\bibfnamefont{E.~L.} \bibnamefont{Pollock}} \bibnamefont{and}
  \bibinfo{author}{\bibfnamefont{D.~M.} \bibnamefont{Ceperley}},
  \bibinfo{journal}{Phys. Rev. B} \textbf{\bibinfo{volume}{36}},
  \bibinfo{pages}{8343} (\bibinfo{year}{1987}).

\bibitem[{\citenamefont{Fisher et~al.}(1989)\citenamefont{Fisher, Weichman,
  Grinstein, and Fisher}}]{fisher1989}
\bibinfo{author}{\bibfnamefont{M.~P.~A.} \bibnamefont{Fisher}},
  \bibinfo{author}{\bibfnamefont{P.~B.} \bibnamefont{Weichman}},
  \bibinfo{author}{\bibfnamefont{G.}~\bibnamefont{Grinstein}},
  \bibnamefont{and} \bibinfo{author}{\bibfnamefont{D.~S.}
  \bibnamefont{Fisher}}, \bibinfo{journal}{Phys. Rev. B}
  \textbf{\bibinfo{volume}{40}}, \bibinfo{pages}{546} (\bibinfo{year}{1989}).

\bibitem[{\citenamefont{Kato and Kawashima}(2010)}]{kato2010}
\bibinfo{author}{\bibfnamefont{Y.}~\bibnamefont{Kato}} \bibnamefont{and}
  \bibinfo{author}{\bibfnamefont{N.}~\bibnamefont{Kawashima}},
  \bibinfo{journal}{Phys. Rev. E} \textbf{\bibinfo{volume}{81}},
  \bibinfo{pages}{011123} (\bibinfo{year}{2010}).

\bibitem[{\citenamefont{Capogrosso-Sansone
  et~al.}(2007)\citenamefont{Capogrosso-Sansone, Prokof'ev, and
  Svistunov}}]{capogrosso2007}
\bibinfo{author}{\bibfnamefont{B.}~\bibnamefont{Capogrosso-Sansone}},
  \bibinfo{author}{\bibfnamefont{N.~V.} \bibnamefont{Prokof'ev}},
  \bibnamefont{and} \bibinfo{author}{\bibfnamefont{B.~V.}
  \bibnamefont{Svistunov}}, \bibinfo{journal}{Phys. Rev. B}
  \textbf{\bibinfo{volume}{75}}, \bibinfo{pages}{134302}
  (\bibinfo{year}{2007}).

\bibitem[{\citenamefont{Capogrosso-Sansone
  et~al.}(2008)\citenamefont{Capogrosso-Sansone, S$\ddot{{\rm o}}$yler,
  Prokof'ev, and Svistunov}}]{capogrosso2008}
\bibinfo{author}{\bibfnamefont{B.}~\bibnamefont{Capogrosso-Sansone}},
  \bibinfo{author}{\bibfnamefont{S.~G.} \bibnamefont{S$\ddot{{\rm o}}$yler}},
  \bibinfo{author}{\bibfnamefont{N.}~\bibnamefont{Prokof'ev}},
  \bibnamefont{and}
  \bibinfo{author}{\bibfnamefont{B.}~\bibnamefont{Svistunov}},
  \bibinfo{journal}{Phys. Rev. A} \textbf{\bibinfo{volume}{77}},
  \bibinfo{pages}{015602} (\bibinfo{year}{2008}).

\bibitem[{\citenamefont{Campostrini et~al.}(2001)\citenamefont{Campostrini,
  Hasenbusch, Pelissetto, Rossi, and Vicari}}]{campostrini2001}
\bibinfo{author}{\bibfnamefont{M.}~\bibnamefont{Campostrini}},
  \bibinfo{author}{\bibfnamefont{M.}~\bibnamefont{Hasenbusch}},
  \bibinfo{author}{\bibfnamefont{A.}~\bibnamefont{Pelissetto}},
  \bibinfo{author}{\bibfnamefont{P.}~\bibnamefont{Rossi}}, \bibnamefont{and}
  \bibinfo{author}{\bibfnamefont{E.}~\bibnamefont{Vicari}},
  \bibinfo{journal}{Phys. Rev. B} \textbf{\bibinfo{volume}{63}},
  \bibinfo{pages}{214503} (\bibinfo{year}{2001}).

\bibitem[{\citenamefont{Hasenbusch}(2010)}]{hasenbusch2010}
\bibinfo{author}{\bibfnamefont{M.}~\bibnamefont{Hasenbusch}},
  \bibinfo{journal}{Phys. Rev. B} \textbf{\bibinfo{volume}{82}},
  \bibinfo{pages}{174433} (\bibinfo{year}{2010}).

\bibitem[{\citenamefont{Kosterlitz and Thouless}(1973)}]{kosterlitz1973}
\bibinfo{author}{\bibfnamefont{J.~M.} \bibnamefont{Kosterlitz}}
  \bibnamefont{and} \bibinfo{author}{\bibfnamefont{D.~J.}
  \bibnamefont{Thouless}}, \bibinfo{journal}{J. Phys. C}
  \textbf{\bibinfo{volume}{6}}, \bibinfo{pages}{1181} (\bibinfo{year}{1973}).

\bibitem[{\citenamefont{Kosterlitz}(1974)}]{kosterlitz1974}
\bibinfo{author}{\bibfnamefont{J.~M.} \bibnamefont{Kosterlitz}},
  \bibinfo{journal}{J. Phys. C} \textbf{\bibinfo{volume}{7}},
  \bibinfo{pages}{1046} (\bibinfo{year}{1974}).

\bibitem[{\citenamefont{Weber and Minnhagen}(1988)}]{weber1988}
\bibinfo{author}{\bibfnamefont{H.}~\bibnamefont{Weber}} \bibnamefont{and}
  \bibinfo{author}{\bibfnamefont{P.}~\bibnamefont{Minnhagen}},
  \bibinfo{journal}{Phys. Rev. B} \textbf{\bibinfo{volume}{37}},
  \bibinfo{pages}{5986} (\bibinfo{year}{1988}).

\bibitem[{\citenamefont{Harada and Kawashima}(1998)}]{harada1998}
\bibinfo{author}{\bibfnamefont{K.}~\bibnamefont{Harada}} \bibnamefont{and}
  \bibinfo{author}{\bibfnamefont{N.}~\bibnamefont{Kawashima}},
  \bibinfo{journal}{J. Phys. Soc. Jpn.} \textbf{\bibinfo{volume}{67}},
  \bibinfo{pages}{2768} (\bibinfo{year}{1998}).

\end{thebibliography}

\end{document}